\documentclass[aastex6, revtex4, twocolumn]{emulateapj}
\usepackage{amsmath}
\usepackage{array}
\usepackage{hyperref}
\hypersetup{colorlinks=true, linkcolor=blue, citecolor=blue, urlcolor=blue}
\usepackage{graphicx}
\usepackage{gensymb} 
\usepackage{lscape}
\usepackage{longtable}
\usepackage{rotating}

\shorttitle{Chromospherically Active Stars in the RAVE Survey. II. Young dwarfs in the Solar neighborhood}
\shortauthors{\v Zerjal et al.}

\begin{document}
\title{Chromospherically Active Stars in the RAVE Survey. II. Young dwarfs in the Solar neighborhood}

\author{
M.~\v Zerjal\altaffilmark{1}, 
T.~Zwitter\altaffilmark{1}, 
G.~Matijevi\v c\altaffilmark{2},
E.~K.~Grebel\altaffilmark{3},
G.~Kordopatis\altaffilmark{2},
U.~Munari\altaffilmark{4},
G.~Seabroke\altaffilmark{5},
M.~Steinmetz\altaffilmark{2},
J.~Wojno\altaffilmark{2},
O.~Bienaym\'{e}\altaffilmark{6},
J.~Bland-Hawthorn\altaffilmark{7},
C.~Conrad\altaffilmark{2},
K.~C.~Freeman\altaffilmark{8},
B.~K.~Gibson\altaffilmark{9},
G.~Gilmore\altaffilmark{10},
A.~Kunder\altaffilmark{2},
J.~Navarro\altaffilmark{11},
Q.~A.~Parker\altaffilmark{12,13}, 
W.~Reid\altaffilmark{14,15},
A.~Siviero\altaffilmark{16},
F.~G.~Watson\altaffilmark{17} and
R.~F.~G.~Wyse\altaffilmark{18}
}

\altaffiltext{1}{Faculty of Mathematics and Physics, University of Ljubljana, Jadranska 19, 1000 Ljubljana, Slovenia; \email{marusa.zerjal@fmf.uni-lj.si}}
\altaffiltext{2}{Leibniz-Institut f\"ur Astrophysik Potsdam (AIP), An der Sternwarte 16, D-14482, Potsdam, Germany}
\altaffiltext{3}{Astronomisches Rechen-Institut, Zentrum f\"ur Astronomie der Universit\"at Heidelberg, M\"onchhofstr. 12-14, D-69120 Heidelberg, Germany}
\altaffiltext{4}{INAF Osservatorio Astronomico di Padova, 36012 Asiago, Italy}
\altaffiltext{5}{Mullard Space Science Laboratory, University College London, Holmbury St Mary, Dorking, RH5 6NT, UK}
\altaffiltext{6}{Observatoire astronomique de Strasbourg, Universit\'{e} de Strasbourg, CNRS, UMR 7550, 11 rue de l’Universit\'{e}, F-67000 Strasbourg, France}
\altaffiltext{7}{Sydney Institute for Astronomy, School of Physics, University of Sydney, NSW 2006, Australia}
\altaffiltext{8}{Research School of Astronomy and Astrophysics, Australia National University, Weston Creek, Canberra, ACT 2611, Australia}
\altaffiltext{9}{E.A. Milne Centre for Astrophysics, University of Hull, Hull, HU6 7RX, United Kingdom}
\altaffiltext{10}{Institute of Astronomy, University of Cambridge, Madingley Road, Cambridge CB3 0HA, United Kingdom}
\altaffiltext{11}{Senior CIfAR Fellow, University of Victoria, Victoria, BC, Canada V8P 5C2}
\altaffiltext{12}{Department of Physics, The University of Hong Kong, Hong Kong SAR, China}
\altaffiltext{13}{Laboratory for Space Research, The University of Hong Kong, Hong Kong SAR, China}
\altaffiltext{14}{Department of Physics and Astronomy, Macquarie University, Sydney, NSW 2109, Australia}
\altaffiltext{15}{Western Sydney University, Locked Bag 1797, Penrith South DC, NSW 1797, Australia}
\altaffiltext{16}{Dipartimento di Fisica e Astronomia Galileo Galilei, Universita’ di Padova, Vicolo dell’Osservatorio 3, I-35122 Padova, Italy}
\altaffiltext{17}{Australian Astronomical Observatory, PO Box 915, North Ryde, NSW 1670, Australia}
\altaffiltext{18}{Johns Hopkins University, Homewood Campus, 3400 N Charles Street, Baltimore, MD 21218, USA}



\begin{abstract}
A large sample of over 38,000 chromospherically active candidate solar-like stars and cooler dwarfs from the RAVE survey is addressed in this paper. An improved activity identification with respect to the previous study was introduced to build a catalog of field stars in the Solar neighborhood with an excess emission flux in the calcium infrared triplet wavelength region.

The central result of this work is the calibration of the age--activity relation for the main sequence dwarfs in a range from a few $10 \; \mathrm{Myr}$ up to a few Gyr. It enabled an order of magnitude age estimation of the entire active sample. Almost 15,000 stars are shown to be younger than $1\;\mathrm{Gyr}$ and $\sim$2000 younger than $100\;\mathrm{Myr}$. The young age of the most active stars is confirmed by their position off the main sequence in the $J-K$ versus $N_{UV}-V$ diagram showing strong ultraviolet excess, mid-infrared excess in the $J-K$ versus $W_1-W_2$ diagram and very cool temperatures ($J-K>0.7$). They overlap with the reference pre-main sequence RAVE stars often displaying X-ray emission. The activity level increasing with the color reveals their different nature from the solar-like stars and probably represents an underlying dynamo generating magnetic fields in cool stars. 

50\% of the RAVE objects from DR5 are found in the TGAS catalog and supplemented with accurate parallaxes and proper motions by Gaia. This makes the database of a large number of young stars in a combination with RAVE's radial velocities directly useful as a tracer of the very recent large-scale star formation history in the Solar neighborhood. The data are available online in the Vizier database.
\end{abstract}

\section{Introduction}
A large portion of solar-like and later-type dwarf stars exhibit signs of chromospheric activity during their young ages, especially before they reach Solar age (\citealp{2008ApJ...687.1264M}, hereafter M08).
Chromospheric activity develops in stars with a subsurface convective layer, i.e., stars in the cooler part of the Hertzsprung-Russell diagram. There are two components responsible for the excessive emission flux. The dominant component is produced by a complex rotation-driven magnetic dynamo. It is superimposed on the basal emission (e.g., \citealp{1989ApJ...337..964S}) originating from the acoustic energy released into the atmosphere from the ubiquitous convective cells. While the basal emission remains constant the loss of the angular momentum of the star over time leads to the decline of the rotation period and activity. Because the decline of the excess flux over hundreds of millions of years exceeds the short-term variations on the scales of days (stellar flares), months (starspot rotation) and decades (Solar 11-year cycle analogues) - e.g., \citealp{2004ApJS..152..261W} - this observable is a suitable age estimator. While precise dating is not possible due to many reasons (e.g. pre-main sequence and zero-age main sequence stars exhibit a broad range of activity levels at a given age, \citealp{2010ARA&A..48..581S}), 
an order of magnitude age estimate is easily attainable in the range from a few tens of millions of years up to a few Gyr.

In contrast to the more fundamental but observationally demanding gyrochronology-based dating, the detection of chromospheric activity is straightforward as it manifests itself in a wide range of emission intensities in the strongest spectral lines ($\mathrm{H\alpha}$, $\mathrm{H\beta}$, Ca~II~H\&K, Ca~II~infrared triplet; Mg~II~h\&k -- the latter not visible from the ground). A single spectral measurement with moderate signal-to-noise ratio ($>20$ per pixel) and mid-range resolving power is adequate for stellar age estimation using the age-activity relation. This fact is of huge importance in the era of large Milky Way spectroscopic surveys covering hundreds of thousands of stars as it enables the age estimation of a large number of young candidates. 
Such catalogs of young stars enhanced with the individual abundances (e.g., Galah (\citealp{2015MNRAS.449.2604D}) and Gaia-ESO (\citealp{2014A&A...570A.122S}) Surveys) and astrometric data and parallaxes from Gaia (\citealp{2015A&A...574A.115M}) hold a huge potential for the investigation not only of the recent star formation history in the Solar neighborhood and the evolution of the Milky Way galaxy but the nature of stellar dynamo mechanisms and the influence of stellar activity and abundances on the planetary environments as well.



In this paper we investigate the data from the RAVE Survey (RAdial Velocity Experiment\footnote{https://www.rave-survey.org/}, \citealp{2006AJ....132.1645S, 2008AJ....136..421Z, 2011AJ....141..187S, 2013AJ....146..134K, 2016arXiv160903210K}). While most of the literature studying chromospheric activity covers the strong Ca~II~H\&K (3967 and $3933\;\mathrm{\AA}$; \citealp{2015RAA....15.1282Z} for example report on a catalog of 120,000 F, G, and K stars with the Ca~II~H\&K emission in the first LAMOST data release), RAVE focuses on the Ca~II infrared triplet (Ca~II~IRT; 8498, 8542 and $8662\;\mathrm{\AA}$). The latest data release (DR5, \citealp{2016arXiv160903210K}) includes 520,781 multi-fibre spectroscopic measurements of 457,588 stars collected between 2003 and 2013 with radial velocity determination being one of the main goals of the survey.

With the help of an efficient unsupervised stellar classification algorithm based on normalized stellar fluxes and independent of stellar parameters (locally linear embedding, \citealp{2012ApJS..200...14M}, hereafter M12) tens of thousands of candidate active stars have been uncovered. 
A large catalog of 44,000 candidate chromospherically active RAVE stars described in the first paper in the series (\citealp{2013ApJ...776..127Z}, hereafter Z13) contains over 14,000 stars above the $2\sigma$ excess emission detection limit. 

Around 65\% of all the objects in the RAVE database (and 60\% of the active candidates) are also found in the Tycho-2 astrometric catalog (\citealp{2000A&A...355L..27H}). Because these dwarfs reside in the Solar neighborhood their distance errors in the first Gaia data release (TGAS: ``using the positions from the Tycho-2 Catalog as additional information for a joint solution with early Gaia data"; \citealp{2015A&A...574A.115M}) are expected to be of the order of 10\%, a number that will become considerably more precise with future data releases (reaching a level of $\sim 1$\%, e.g., \citealp{2013A&A...559A..74B}).
The catalog of active stars in combination with radial velocities (provided by RAVE) and reliable distances and proper motions will directly enable the study of the early stellar evolution and recent star formation history in the local neighborhood.

The dataset will be significantly enhanced by the Gaia-RVS catalog (Radial Velocity Spectrometer, covering the same Ca~II~IRT domain, \citealp{2004MNRAS.354.1223K,2011EAS....45..181C}). By the end of the mission the set of around 9~million stars (with their $G_{RVS}$ magnitude brighter than $\sim 11.75$ and the signal-to-noise ratio of their spectra over 20 per pixel) will be the largest database to look for active stars in.

In paper I (Z13) we listed our selection of candidate active stars, in the present paper II we derive their ages. First we introduce an improved activity identification procedure by elimination of stars with non-chromospheric sources of emission from the original catalog (Sec. \ref{sec.data}). The catalog is enhanced with 46 additional very active candidates not included in Z13. A study of the $J-K$ versus $N_\mathrm{UV}-V$ diagram in Sec. \ref{sec.photometry} shows an offset of the youngest stars with the highest emission levels from the main sequence. This group of stars accumulates in a relatively confined part of the plot where young, T~Tauri, pre-main sequence stars and stars with X-ray emission are found. A calibration of the age--activity relation using ages from the literature is presented in Sec. \ref{sec.age} as the leading result of this paper which allows an order of magnitude age determination for more than 22,000 young RAVE stars in the Solar neighborhood. The lower and upper distance limit determination is described in Sec. \ref{sec.distance}.
Conclusions with a discussion, open questions and future plans including orbital simulations of the youngest stars using Gaia astrometry are presented in Sec. \ref{sec.discussion}. A table of the external reference ages used in the age--activity relation calibration is presented in the Appendix.

\section{Improved catalog} \label{sec.data}
The catalog of candidate active stars used in this work is described in detail by Z13. Here we very briefly review the determination of activity levels and basic characteristics of the sample to allow the reader to better understand the context of present study.


Chromospheric activity in the RAVE domain manifests itself in an excessive emission flux in the Ca~II~IRT while the rest of the spectrum remains indistinguishable from an inactive state. The strength of the emission can range from marginally detectable levels to individual cases with fluxes above the continuum level.
The qualitative classification of the RAVE database using the locally linear embedding (LLE) technique (M12) entirely based on normalized spectra revealed over 44,000 stars with the possible presence of an excessive emission in the Ca~II~IRT. For the detailed selection criteria see Z13.

The photospheric component of the flux was eliminated from the active candidate star by the subtraction of the best-matching inactive template spectrum.
For the purpose of the spectral subtraction technique a database of over 12,000 inactive single solar-like or later-type main sequence dwarfs with no emission or any other peculiarities was built from the measured RAVE sample.
Although chromospheric activity has been detected in evolved stars (e.g. \citealp{1976ApJ...205..823W}), objects with $\log{g}<3.5$ were excluded from the inactive sample because magnetic activity is not expected to be notably present in giant stars above the main sequence. \citealp{1984A&A...130..353R} (hereafter R84) showed that the emission rate in evolved stars is very close to the basal level.

The number of stars in the inactive set is sufficient to cover the entire parameter space of active candidates, including the effective temperature, gravity and metallicity, varied noise realizations as well as possible variations of resolving power along the spectra. Most of the inactive stars are concentrated around the Solar temperature. It becomes less likely to find an inactive star in the cooler regions of the main sequence because active stars become more and more dominant there.
Due to lack of inactive stars above the main-sequence the same database of 12,000 stars is used for the activity estimation of the pre-main sequence stars.



\begin{figure}
\includegraphics[width=\columnwidth]{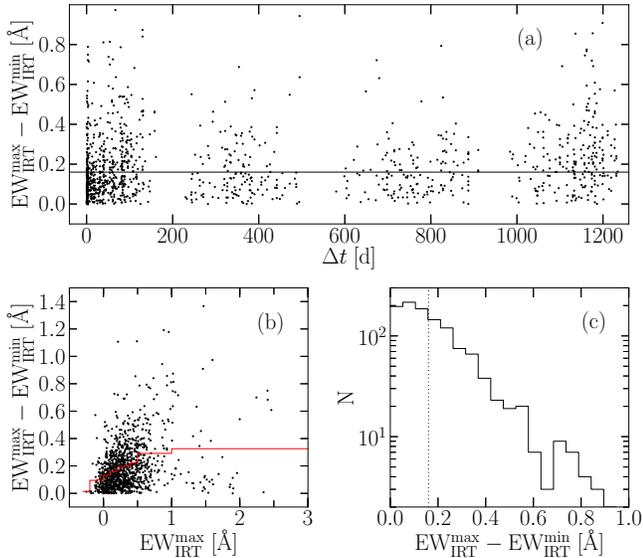}
\caption{Time variability of the Ca~II~IRT chromospheric emission for stars with multiple measurements measured as the difference between the maximal and minimal activity rate. Activity detection limit ($0.16\;\mathrm{\AA}$) is shown with the black line (a). The majority of stars in the plot have 2 repeated observations, $\Delta t$ is a time difference between the measurements. More active stars exhibit higher variability rates (panel b; the moving average is shown with red line: bin size between -0.5 and $0.5\;\mathrm{\AA}$ is $0.1\;\mathrm{\AA}$ and larger above $0.5\;\mathrm{\AA}$). The median variability of the sample ($0.15\;\mathrm{\AA}$) is comparable with the activity detection limit (c).}
\label{fig.time_variability}
\end{figure}

The advantage of the approach using the measured inactive library from the RAVE Survey in contrast to a synthetic dataset is the fact that active and inactive sets share the same instrumental profile of the spectra (e.g. resolving power and point-spread function along the spectra). It avoids problems with an invalid assumption of the local thermodynamic equilibrium in the chromospheric layer used in models which affects the cores of the strongest lines that are at the same time sensitive to magnetic activity. Moreover, the determination of the best-matching templates is parameter-free because it is based solely on the comparison of the normalized fluxes of active and inactive spectra. Atmospheric parameters are excluded from the template search. This is an important aspect as values for the highly active stars, together with the radial velocities, could be inaccurate due to the influence of the excessive flux. Additionally, the parameter estimation for RAVE stars with a temperature cooler than 4000~K shows systematic offsets. The search for the best-matching template by a comparison of the normalized fluxes was performed in an iterative algorithm to apply radial velocity corrections. To avoid the impact of the shallower calcium lines these regions were eliminated across a range of $\pm 2.5 \mathrm{\AA}$ from the line cores of all active and template spectra. Equivalent widths of the excessive emission for each calcium line in the spectrum combined into the sum $\mathrm{EW_{IRT}=EW_{8498}+EW_{8542}+EW_{8662}}$ are used as the activity proxy.


For reference, activity of the inactive database was estimated using the same method. The distribution of equivalent widths in the latter case is centred at $-0.05\;\mathrm{\AA}$ with a standard deviation of $\sigma=0.16\;\mathrm{\AA}$. An investigation of the subtracted spectra revealed that their average value outside calcium lies slightly below zero ($\sim 10^{-3}$ in the normalized flux units) which translates to the negative offset of the emission equivalent width. $\sigma$ is accepted as an internal error on the derived $\mathrm{EW_{IRT}}$. 

Another contribution to the uncertainty is the intrinsic time variability of the activity on the scales of days (stellar flares), months (starspot rotation) and decades (Solar 11-year cycle analogues). The statistics of the repeated observations of the same active stars from the active RAVE catalog (1146 stars, of which 30\% were observed more than twice) shows that
the median difference between the minimum and maximum values of activity for a single star ($\mathrm{EW_{IRT}^{max}-EW_{IRT}^{min}}$) is $0.15 \; \mathrm{\AA}$ (Fig. \ref{fig.time_variability}, panel a). In general more active stars show higher variability rates (\citealp{1998ApJS..118..239R}). A moving average of $\mathrm{EW_{IRT}^{max}-EW_{IRT}^{min}}$ versus $\mathrm{EW_{IRT}^{max}}$ confirms the statement (panel b). For stars with $\mathrm{EW_{IRT}^{max}}$ between 1 and $3\;\mathrm{\AA}$ the mean variability is $0.33\;\mathrm{\AA}$ (panel c).
It is important to stress that intrinsic variations of the chromospheric activity occur on time scales from days to months and tens of years (due to flares, stellar rotation of active regions and spots and 11-year Solar cycle analogs). The variations are smaller than the global decline of the activity with time. 


\subsection{The Aquarius overdensity} \label{sec.aqr}
\begin{figure}
\includegraphics[width=\columnwidth]{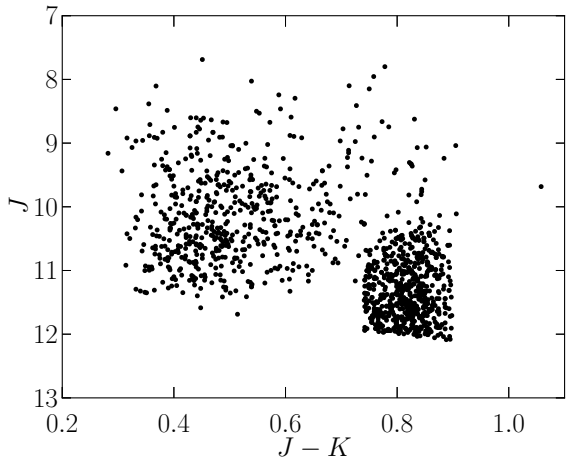} 
\caption{A color-magnitude diagram for Aquarius fields with an overdensity of red stars. J is an apparent magnitude. More red stars were observed to gather more potential Aquarius stream members. Only active candidates are plotted.}
\label{fig.aquarius_cmd}
\end{figure}

The Aquarius stream, which is known to consist of at least 15 stars between $30\degree < l < 75\degree$ and $-70\degree < b < -50\degree$ has been identified in the RAVE data by \citealp{2011ApJ...728..102W}. The stream members are giants at distances of $0.5-10\;\mathrm{kpc}$ with radial velocities $\sim -200\;\mathrm{km\,s^{-1}}$ and estimated to be $10\;\mathrm{Gyr}$ old.
An additional group of 11 adjacent fields of red stars with $J-K>0.7$ was intentionally observed by RAVE in the direction of the Aquarius constellation at $35\degree<l<75\degree$ and $-62\degree<l<-50\degree$ in order to collect more stream candidates. As a consequence, a supplementary set of red dwarfs was observed as well. As the probability for activity increases toward the lower part of the main sequence, many of them turned out to be active (Fig. \ref{fig.aquarius_cmd}).
Nevertheless, the number of all observed stars in Aquarius and the number of all dwarfs with $\log{g}>3.75$, proper motions and radial velocities show no other peculiarities with respect to similar regions at the same Galactic latitude.

The ratio between red ($J-K\ge 0.7$) active stars and all red RAVE dwarfs in Aquarius is at least 80\%. The same is true for stars with $\mathrm{EW_{IRT}} \ge 1\;\mathrm{\AA}$. 
For this reason the Aquarius fields notably contribute to the statistics of the reddest active stars which are otherwise underpopulated in the RAVE sample due to their low luminosities.




\subsection{Improved activity identification}
\begin{figure*}
\includegraphics[width=\textwidth]{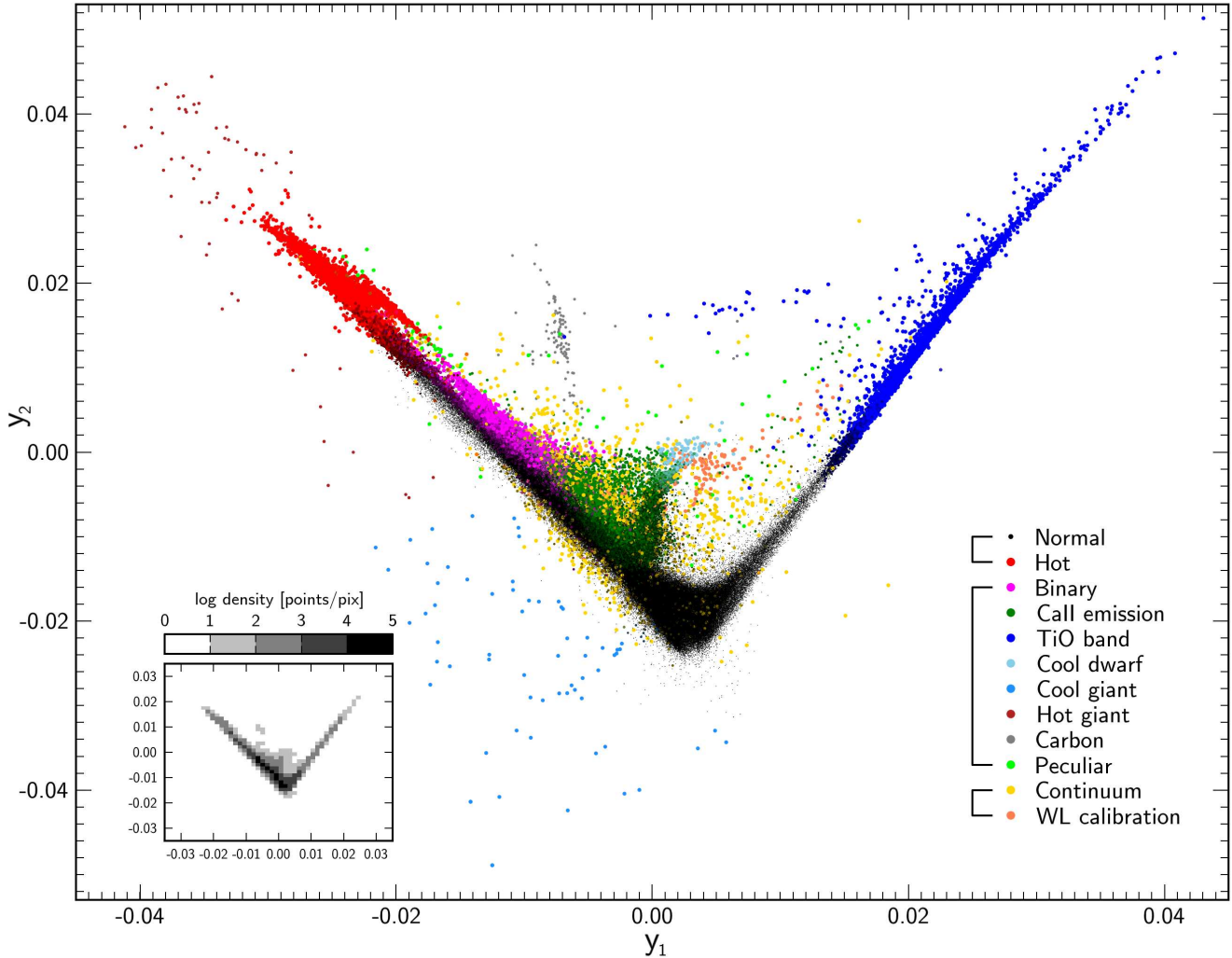} 
\caption{The first two dimensions $y_1$ and $y_2$ of the projection map of the locally linear embedding (LLE) classification technique that conserves relations between the neighboring points of the high-dimensional manifold. A selected spectrum in the projected space is surrounded by its neighbors from the high-dimensional space. Different colors indicate 11 distinct classes of spectra. Stars showing no peculiarities ('Normal stars' in the plot) are intentionally represented with smaller symbols for clarity.  The inset shows the log density of points on the main diagram. Figure adapted from \citealp{2012ApJS..200...14M}.}
\label{fig.lle}
\end{figure*}

The selection of candidates in the catalog is based on the set of morphological flags produced by the spectral classification technique. The classification algorithm compares the flux of the spectrum in question with fluxes from the reference database. 
The reference database was set up in the process involving a locally linear embedding (LLE). LLE is a general dimensionality reduction procedure that conserves relations between the neighboring points of the high-dimensional manifold. Because a selected spectrum in the projected space is surrounded by its neighbors from the high-dimensional space, the algorithm is useful for the classification purposes. An example of the projection of the RAVE data in the two-dimensional space is shown in Fig. \ref{fig.lle} where different colors indicate 11 distinct classes of spectra.

The method includes three steps. After the nearest neighbors are found for each data point in the original space, the set of weights that best describe the data point as a linear combination of its neighbors is derived. The final, key step is a projection on the low-dimensional space where each point is still best represented by the same weights from the previous step. For the details about the key steps see M12 or \citet{2000Science...190...2323R, 2009AJ....138.1365V} and references therein.

The classification of the RAVE data was broken down into two crucial steps. First, a reference database of the most representative $\sim 5000$ spectra was established by iteratively sifting the most dense areas of the LLE projection containing the most populated groups of stars. This way all distinct classes of objects were represented as evenly as possible in the final dataset. The reference set was then manually assigned classification flags of 11 distinct morphological classes. The classification of the rest of the RAVE spectra was calculated in the second step by comparing the spectra to the reference set. The outcome for each spectrum is a list of 20 flags ordered by the distance between the flux and its nearest reference neighbors. For more details see M12.

A star was selected as an active candidate when at least one nearest neighbor within its first 6 neighbors from the LLE reference database showed an emission-type spectrum. While this relatively loose criterion helps to unravel marginally active spectra, it on the other hand contaminates the catalog with individual peculiar stars (where some of their nearest neighbors show unsuitable morphology, e.g., binaries, giants etc.). 
The aim of an improved activity identification is thus the elimination of active candidates that show peculiarity other than an excess emission in the Ca~II~IRT.

The search for such stars was performed in a few independent ways including nearest neighbor investigation, morphology notes and spectral types from the Simbad database and visual spectral inspection of the most active stars with manual elimination besides manual exclusion of stars with inadequate atmospheric parameters. Each star was examined by all these criteria.

The largest group of suspicious active candidates were giants with $\log{g}<3.7$ (6147 stars). 
Such stars were excluded from the candidate list introduced in Z13.

An additional 68 hot stars with Paschen lines and shallow calcium lines mistaken for activity were recognized manually.
If any of the repeated observations (where available) were recognized as binary all spectra of such objects were removed (101 spectra).

Spectra with $\mathrm{EW_{IRT}>1.7\AA}$ were manually checked for systematic errors. A few additional cases with peculiar fluxes were eliminated.

\begin{table}
\centering
\caption{Number of stars eliminated from the initial active candidate catalog for each type of peculiarity.}
\begin{tabular}{l l}
\hline
\hline
Type of peculiarity & N \\
\hline
$\log{g}<3.7$ & 5465 \\
LLE b & 1490 \\
Double or Binary & 541+37 \\
LLE e, $\log{g}<3.7$ & 323 \\
LLE e & 194 \\
SimbadSpType & 110 \\
LLE c & 107 \\
LLE b, $\log{g}<3.7$ & 104 \\
$\log{g}<3.7$, SimbadSpType & 59 \\
LLE e, LLE o & 56 \\
LLE b, Binary or Double & 51+34 \\
LLE e, LLE b & 46 \\
$\log{g}<3.7$, Double & 44 \\
Galaxy & 34 \\
LLE e, LLE o, Hot & 28 \\
$\log{g}<3.7$, LLE t, LLE p, $J-K>0.6$ & 27 \\
LLE e, SimbadSpType & 26 \\
LLE c, $\log{g}<3.7$ & 22 \\
LLE e, $\log{g}<3.7$, SimbadSpType & 21 \\
Mira & 18 \\
LLE e, $\log{g}<3.7$, LLE o & 14 \\
LLE b, SimbadSpType & 12 \\
LLE e, LLE o, SimbadSpType & 12 \\
LLE t, LLE p, $J-K>0.6$ & 11 \\
BY Dra & 10 \\
Other & 242 \\
\hline
Total & 9138 \\
\hline
\end{tabular}
\label{tab.peculiar_stars_elimination_numbers}
\end{table}

The nearest LLE neighbor investigation revealed binary candidates (marked as LLE~b) with more than 2 binary neighbors (or the first or the second neighbor is a binary), stars with continuum problems (LLE~c), wavelength calibration errors (LLE~w), hot stars (LLE~o), stars with TiO bands (LLE~t) and other types of peculiarity (LLE~p).

LLE~e denotes stars with no emission-type neighbors (a few 100 stars). These stars were originally included in the catalog because they are candidate members of young clusters and were tested for possibly overlooked emission, especially for cases with a faulty radial velocity shift of several $100\;\mathrm{km\,s^{-1}}$ due to very strong emission. They were removed from the sample because they do not meet the LLE selection criteria.

A cross-check using the online Simbad database for stars with available morphology types (25,238 objects) confirmed the presence of binaries, Miras (marked as 'LLE t, LLE p, $J-K>0.6$' in the table), Cepheids, galaxies (34 cases; most objects are actually stars with galaxies located nearby within $1\;\mathrm{arcsec}$), stars of BY~Dra, RS~CVn and RR~Lyr type, 9 (micro)lensing events, red giant branch and carbon stars, gamma-ray sources and 2 stars of an unknown nature. The frequency of peculiar Simbad morphology types is less than 3\%.
Note that spectra of RS~CVn-type stars can look identical to active single stars and BY~Dra stars can mimic chromospheric activity. Because the frequency of such known stars in the active sample is relatively small (13 RS~CVn and 28 BY~Dra objects) it is assumed that only a few individual cases could be left unrecognised in the database.

Simbad spectral types ('SimbadSpType') available for 4691 stars uncovered B, A and F-types and giants (luminosity classes I, II and III). The rest of the spectra has a clear peak on the main sequence: most of the stars are G- and K-type dwarfs.

\begin{figure*}
\includegraphics[width=\textwidth]{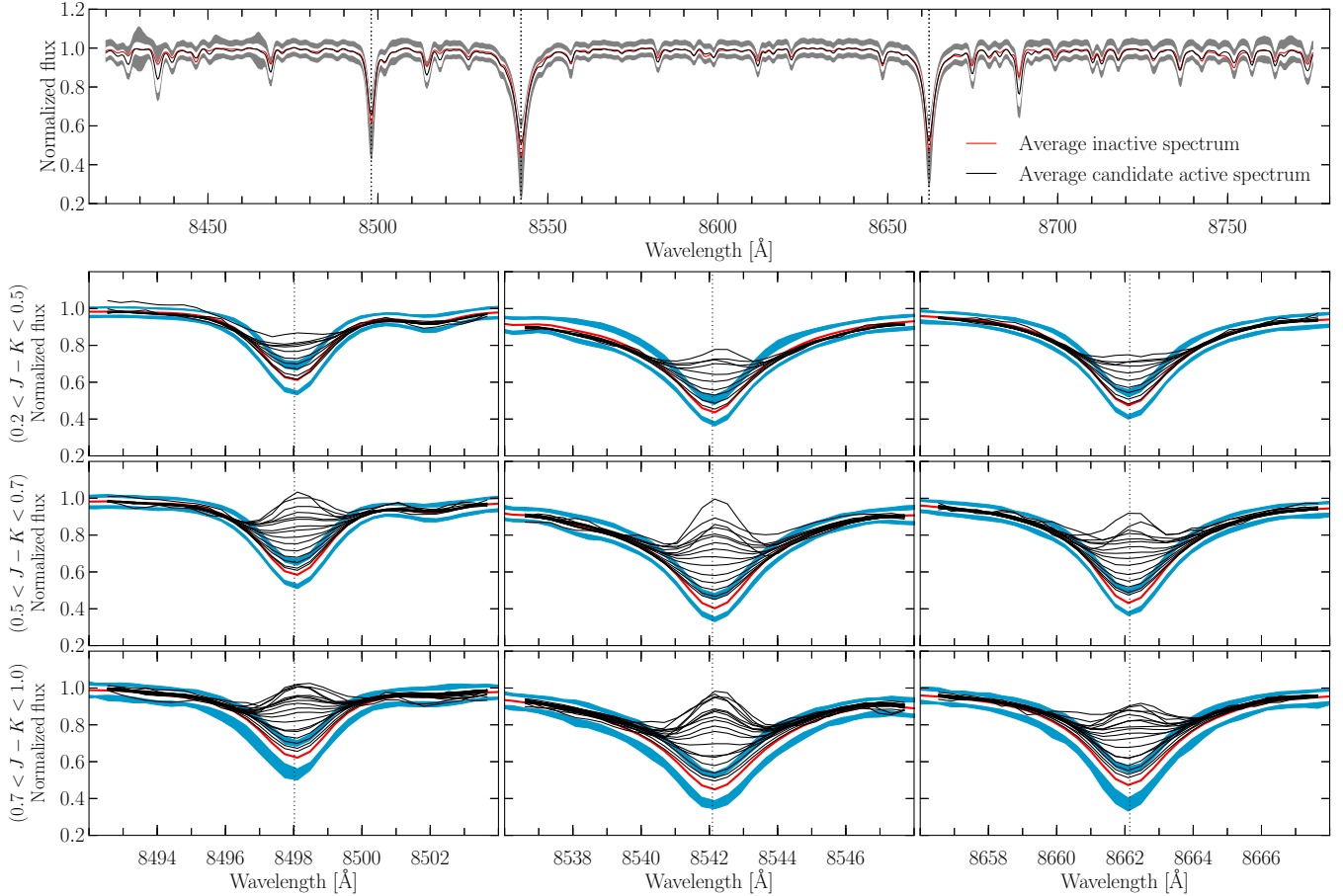} 
\caption{\textit{Top}: The averaged active spectrum ($\mathrm{EW_{IRT}>0.1\;\AA}$) (black line) together with its $1\sigma$ (white) and $2\sigma$ (grey) standard deviation. Averaged inactive spectrum is shown for comparison (red line). Except for the TiO band at $\sim 8450\;\mathrm{\AA}$ and the strongest lines (calcium excess flux, Ti~I ($8434.98\mathrm{\AA}$), Fe~I ($8688.63\mathrm{\AA}$) etc.) the spectra remain unchanged because the number of marginally active spectra in the active sample is large.
\textit{Bottom three rows:} Focus on calcium lines for averaged spectra for each of the increasing activity bins (a range from 0 to $3.6\;\mathrm{\AA}$ with a stepsize of $0.2\;\mathrm{\AA}$) divided into three color classes. For bins below $0.8\;\mathrm{\AA}$ 1000 spectra are averaged. The number drops for more active spectra from few hundreds ($<2.2\;\mathrm{\AA}$), few tens ($<2.8\;\mathrm{\AA}$) to less than 10 for the most active bins. White and blue bands indicate 1 and $2\sigma$ standard deviations for inactive spectra and include parameter variations (metallicity and temperature to some extent within a selected color bin). Because each active spectrum was compared to its nearest neighbor (and not to an averaged inactive spectrum) the fact that the least active spectra lie within 1 or $2\sigma$ bands does not necessarily mean that they do not differ from their inactive counterpart. For the most active spectra line asymmetries (towards red) appear.}
\label{fig.average_spectrum}
\end{figure*}




9138 stars were removed in total from the initial catalog. A list of the number of stars meeting either of the rejection criteria described above is given in Table \ref{tab.peculiar_stars_elimination_numbers}.


46 new active (young) candidates were found during the Simbad classification study (`T~Tau', `Pre-main sequence' and `Young' types). These stars were not recognized as active by the LLE technique because they do not show signs of activity or have large radial velocity shifts (not derived correctly due to their strong emission) that make them look very peculiar.
However, when allowing for these radial velocity shifts, 24 stars out of 46 were found to have $\mathrm{EW_{IRT}>0.1\;\AA}$ and 16 with $\mathrm{EW_{IRT}>1\;\AA}$.

Inspection of the new spectra, first listed in DR5, revealed new active candidates. After the removal of unsuitable cases following the same requirements as above, 2919 new spectra of 2882 stars were found to show signs of activity. The sample was added to the catalog.

\bigskip

An averaged spectrum in the decontaminated catalog together with 1 and $2\;\mathrm{\sigma}$ flux variations is compared with the averaged spectrum from the inactive database in Figure \ref{fig.average_spectrum}. The main contribution to the variations is a range of different atmospheric temperatures (see Sec. \ref{sec.photometry} for further discussion) and noise. The average $1\;\mathrm{\sigma}$ variation of the continuum parts is $\sim 0.02$ in the middle and $\sim 0.03$ at the very ends of the spectrum. The value is consistent with the signal-to-noise estimation. Notable variation occurs within the strong lines, e.g. Ti~I $8434.98\;\mathrm{\AA}$, Fe~I $8688.63\;\mathrm{\AA}$, etc. due to the temperature variations of the spectra (and to a smaller extent due to metallicity).

Because marginally active stars outnumber the stars with higher emission levels due to the selection criteria the average calcium lines almost match the inactive profiles. However, averaged spectra within the selected activity ranges (an $\mathrm{EW_{IRT}}$ range from 0 to $3.6\;\mathrm{\AA}$ with stepsize of $0.2\;\mathrm{\AA}$) reveal a continuous and solid increase of excess emission fluxes (Fig. \ref{fig.average_spectrum}, bottom panels), which establishes confidence in the reliability of the data. The most active averaged spectra suffer from more noise because only very few stars fall within these ranges (6 stars between $\mathrm{EW_{IRT}=3\;\AA}$ and $3.2\;\mathrm{\AA}$, 5 stars between $\mathrm{EW_{IRT}=3.2\;\AA}$ and $3.4\;\mathrm{\AA}$ and 2 stars between $\mathrm{EW_{IRT}=3.4\;\AA}$ and $3.6\;\mathrm{\AA}$.)

Besides calcium lines, a mismatch between the averaged inactive and active spectrum occurs in other strong lines because the active database consists of cooler stars than the inactive dataset (see Sec. \ref{sec.photometry} for the comparison).


\begin{table}
\centering
\caption{A comparison between the number of active candidate stars above selected $\sigma$ values and the expected number for the unimodal Gaussian distribution.}
\begin{tabular}{l l | ll | ll}
\hline
\hline
Above $\sigma$ & $p_{\mathrm{log}}$ & $N$ & $\mathrm{Fraction}$ & $\mathrm{Fraction_{exp.}}$ & $N/N_\mathrm{exp.}$\\
\hline
$<1\sigma$ & 0.62 & 16753 & 0.433 & 0.841 & 0.515 \\
$>1\sigma$ & 0.62 & 21925 & 0.567 & 0.159 & 3.7 \\
$>2\sigma$ & 1.10 & 12768 & 0.330 & 0.0228 & 14.5 \\
$>3\sigma$ & 1.77 & 7000 & 0.180 & 0.001 & 134 \\
$>4\sigma$ & 2.63 & 3881 & 0.100 & $3.17 \cdot 10^{-5}$ & 3168 \\
$>5\sigma$ & 3.69 & 2456 & 0.063 & $2.867 \cdot 10^{-7}$ & 221,519 \\
\hline
\end{tabular}
\raggedright{Note: A comparison between the number $N$ of active candidate stars above selected $\sigma$ values (probability $p_{\mathrm{log}}$ is given as well) and the expected number $N_\mathrm{exp.}$ for the unimodal Gaussian distribution with the average of $-0.05\;\mathrm{\AA}$ and $\sigma=0.16\;\mathrm{\AA}$ (these are parameters for the inactive database activity distribution). The ratio $N/N_\mathrm{exp.}$ shows that the number of active stars largely exceeds the expected number from the unimodal distribution. This is quantitative evidence for the bimodality of the distribution of active stars.}
\label{tab.ewirt_above_sigma}
\end{table}



\begin{figure*}
\includegraphics[width=\textwidth]{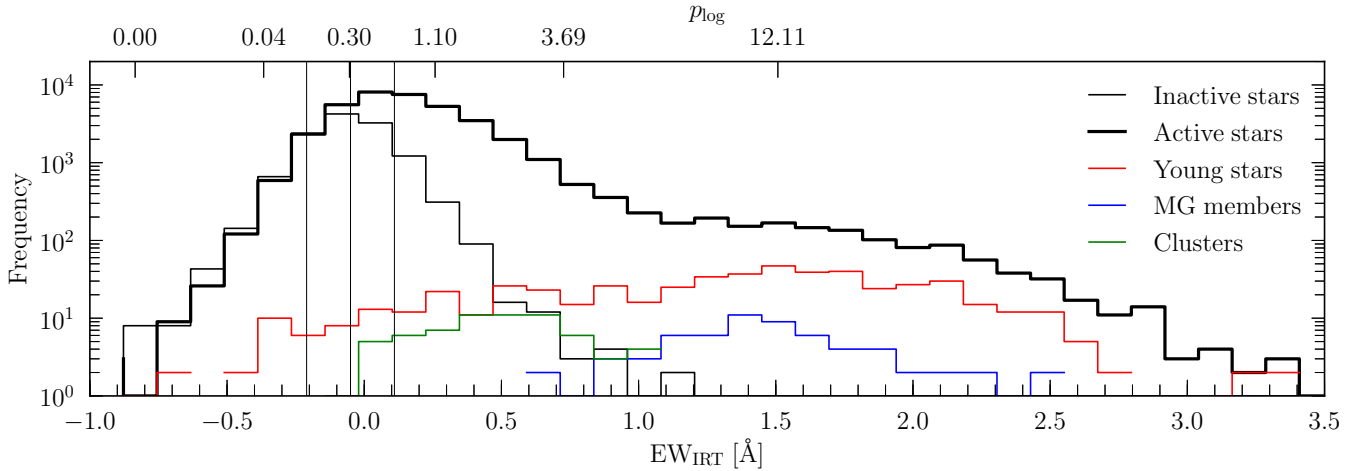} 
\caption{The distribution of activity for active candidates is bimodal (thick black line). The lower end of the less active peak overlaps with the distribution of inactive stars (thin black line) but shows a significant asymmetry with an excess of more active stars. The position of the second peak matches the distribution of young stars (red line) with Simbad morphology types 'Pre-main sequence', 'X-ray source', 'Young Object' or 'T~Tau' star and the position of moving groups (MG; blue line). Cluster members are shown with the green line. Membership of MGs and clusters are listed in Table \ref{tab.reference_stars2}. The bin size is $0.12\;\mathrm{\AA}$. The vertical lines show center and $\pm1\sigma$ deviation of the distribution of inactive stars. The upper abscissa shows the probability that a star with a given activity rate and its uncertainty is more active than an inactive star (for details see Z13).} 
\label{fig.ewirt_distribution}
\end{figure*}

\bigskip
Our new improved catalog of active candidates includes 38,678 spectra of 35,750 stars. Within the sample, the emission levels of 21,925 stars exceed $\mathrm{EW_{IRT}}=0.1\;\mathrm{\AA}$ ($1\,\sigma$) and the activity of almost 13,000 stars surpasses the $2\,\sigma$ emission detection level. The distribution of activity is clearly bimodal (Fig. \ref{fig.ewirt_distribution}). 
A question arises whether the more active peak at $\sim 1.5 \;\mathrm{\AA}$ is an evident feature or simply a tail of the inactive distribution enhanced by the logarithmic scale on the plot. A comparison of the expected fraction of stars (a single Gaussian distribution centered at $-0.05\;\mathrm{\AA}$ and with $\sigma=0.16\;\mathrm{\AA}$ that corresponds to the inactive sample) and the actual fraction of stars (Table \ref{tab.ewirt_above_sigma}) strongly supports the hypothesis of the bimodal distribution. The less active peak is composed of stars with undetectable emission and a moderately active group of stars above $\mathrm{EW_{IRT}}=0.1\;\mathrm{\AA}$. Its skewness does not originate in the imposed tail of the more active peak but stands on its own: the amplitude of the more active peak is $\sim 100$ but at $\mathrm{EW_{IRT}}=0.4\;\mathrm{\AA}$ there are a few thousand more stars than expected. The existence of bimodality is already known in the literature for the Ca~II~H\&K lines (e.g., \citealp{1996AJ....111..439H}).

The more active peak overlaps with the distribution of RAVE stars marked as young, T~Tauri or pre-main sequence types in the Simbad database. Since the active catalog contains a relatively large number of such young stars ($\sim$1600 stars with $\mathrm{EW_{IRT}>1\;\AA}$ which is 4\% of the active sample), the further analysis in this paper is focused on the most active spectra.

\section{Colors of active stars} \label{sec.photometry}
Besides the 2MASS $J$ and $K_\mathrm{s}$ magnitudes (\citealp{2003tmc..book.....C}; 94\% of $J$ magnitude uncertainties and 96\% of $K$ magnitude uncertainties below $0.03\;\mathrm{mag}$ for active RAVE stars) 
additional data were acquired to better characterize the nature of emission objects. The catalog is supplemented by $N_\mathrm{UV}$ fluxes from the GALEX Survey (\citealp{2011MNRAS.411.2770B}; 1771-$2831\;\mathrm{\AA}$; 50\% of stars with uncertainties below $0.04\;\mathrm{mag}$ and 80\% below $0.1\;\mathrm{mag}$), the ROSAT X-ray flux data (0.1-$2.4\;\mathrm{keV}$; \citealp{1999A&A...349..389V}), WISE $W_1$ and $W_2$ mid-infrared bandpasses (3.4 and $4.6\;\mathrm{\mu m}$; \citealp{2014yCat.2328....0C}; typical $W_1$ uncertainties are 0.023 and typical $W_2$ uncertainties are 0.020) and Landolt $V$ magnitudes from the APASS database (typical uncertainties $0.02\;\mathrm{mag}$; \citealp{2014CoSka..43..518H}).

A cross-match between RAVE and the GALEX and the WISE catalogs was performed by the X-match online application (criterion of $5\;\mathrm{arcsec}$ radius).
A manual coordinate match was used for the APASS catalog with $3\;\mathrm{arcsec}$ as a maximum distance between APASS and RAVE stars (the typical APASS astrometric uncertainty is $0.17\;\mathrm{arcsec}$, the fibre size in RAVE is $6.7\;\mathrm{arcsec}$).
X-ray sources were identified via ROSAT 1RXS names obtained through the Simbad online database. It is likely that many of the sources were overlooked but a direct coordinate match was not reliable due to large positional errors in the ROSAT data ($10\;\mathrm{arcsec}$).

\bigskip

\begin{figure}
\includegraphics[width=\columnwidth]{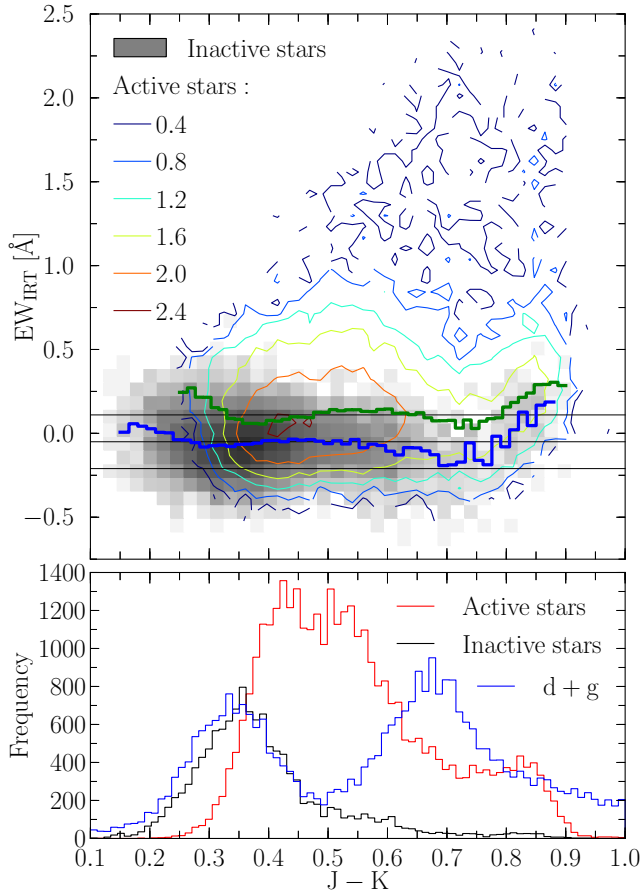} 
\caption{\textit{Top}: Activity versus $J-K$ color for active (contours; the number of stars per bin) and inactive stars (grey shades). Both scales are logarithmic. The highest activity rates are most likely to occur in stars redder than the Sun ($(J-K)_\odot=0.36$). Stars above $J-K>0.7$ show a correlation of activity rate with color (the thick blue line represents the average activity of active stars and the thick green the average activity of inactive stars). Horizontal black lines indicate the center and $\pm 1 \sigma$ deviation of the $\mathrm{EW_{IRT}}$ distribution of inactive stars. 
\textit{Bottom}: Distribution of $J-K$ for active (red line) and inactive stars (black line). Most of the inactive stars ($\sim 12,000$ objects in total) are solar-like while the majority of active stars is cooler (94\% below $5700\;\mathrm{K}$ and 45\% below $5000\;\mathrm{K}$). A slight jump occurs just above $J-K=0.5$ in the active distribution: because at low Galactic latitudes only stars with $J-K>0.5$ were observed the number of red stars is increased. A large number of the coolest (M-type) stars accumulate at $J-K \approx 0.8$. Blue line represents randomly chosen 35,000 RAVE stars (including giants - mostly red clump; histogram values are only qualitative and not to scale with the y axis).}
\label{fig.ewirt_jmk}
\end{figure}

The $J-K$ colors of the majority of active stars lie between the Solar value (0.36) and $\sim0.7$ (Fig. \ref{fig.ewirt_jmk}, lower panel) while the inactive database is solar-like. A bump at $J-K \sim 0.55$ in the color distribution of the active sample originates in the fact that a color-cut was performed in the Galactic plane ($|b|<25\degree$) at $J-K=0.5$ in order to observe more red giants when the RAVE observations were defined. The distribution at $J-K>0.5$ is thus more populated.

A number of red stars concentrates around $J-K\sim 0.8$. The reader may recall the additional observations of the Aquarius stars described in Sec. \ref{sec.aqr}. However, since there are 715 Aquarius red active stars observed out of 6774 red ones in total (11\%), this is not the main contribution to the population. 

\begin{figure*}
\centering
\includegraphics[width=0.8\textwidth]{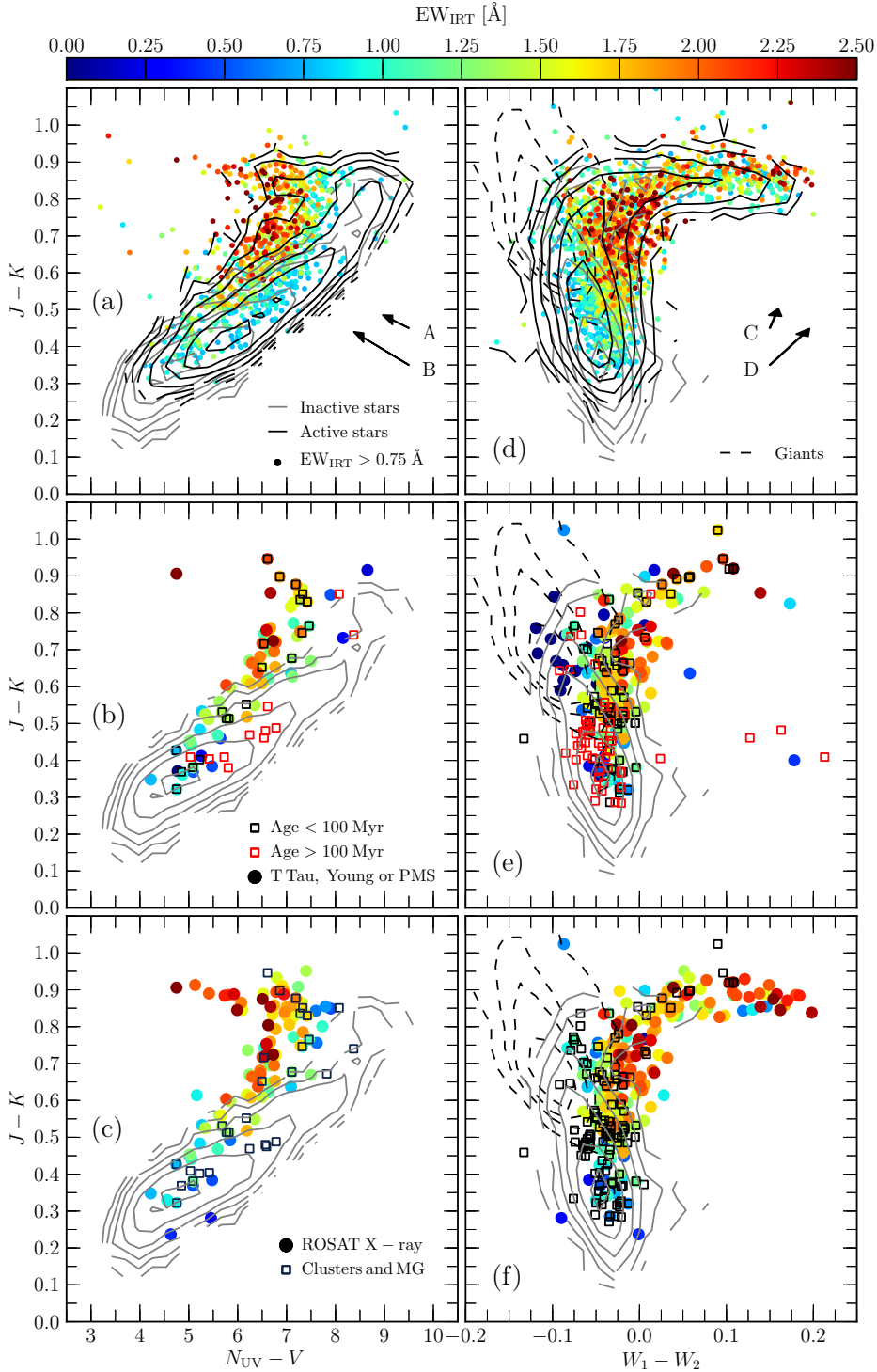} %
\caption{$J-K$ versus $N_\mathrm{UV}-V$ (left) and $W_1-W_2$ (right) diagrams for active stars. Grey contours represent the reference distribution of inactive stars (i.e., the main sequence) in all panels. All contours are plotted in a logarithmic scale.
\textit{Panels a and d}: The distribution of all active candidates (black contours); stars with $\mathrm{EW_{IRT}}>0.75\;\mathrm{\AA}$ are additionally illustrated with dots. Note a concentration of the most active stars at $J-K>0.7$ and $5<N_\mathrm{UV}-V<8$ ($W1-W2 \approx 0$) and the fact that the most active stars in panels \textit{a, b and c} lie off the main sequence. Arrows indicate the average shift in comparison with the colors of the first inactive nearest neighbors of stars for activities between 0.5 and $1\;\mathrm{\AA}$ (A and C) and above $1\;\mathrm{\AA}$ (B and D). The legend in the left panel represents symbols used in both near-ultraviolet and mid-infrared plots. The distribution of giants (dashed contours; only the densest regions are shown) is given for reference.
Panels b and e show young stars (dots; according to Simbad) and stars with known ages (see Table \ref{tab.reference_stars2}) while X-ray sources, cluster and moving group members are depicted in panels c and f. Simbad stars showing reliable indicators of youth overlap with the most active RAVE stars and demonstrate their young ages.
For more details see text.}
\label{fig.photometry_young}
\end{figure*}

The nature of red stars is revealed by the the mid-infrared WISE photometry (Fig. \ref{fig.photometry_young}, panel \textit{d}). Objects with $J-K>0.7$ and $-0.1 \lesssim W_1-W_2 \lesssim 0.2$ are cool (probably M-type) dwarfs. A similar analysis separating cool giants and dwarfs was done by \citealp{2016ApJ...823...59L}. A comparison between the active (black contours) and inactive sample (grey contours) shows a relative lack of cool inactive dwarfs because the majority of stars is active in this region of the color-color diagram. A number of very active stars at $J-K\approx 0.7$ concentrates around $W_1-W_2 \approx 0$. Dashed contours representing giants (30,000 randomly chosen RAVE stars with $\log{\mathrm{g}}<3.5$ and $T_\mathrm{eff}<6000\;\mathrm{K}$) do not overlap with the very active area. This fact, together with the distribution of giants over $J-K $ compared to the distribution of the active sample (Fig. \ref{fig.ewirt_jmk}) excludes the possibility that the very active stars are mistaken for the old red clump giants.

A similar diagram (left column - panels a, b and c) is shown using the near-ultraviolet photometry ($N_\mathrm{UV}-V$ colors).
In both cases (near-ultraviolet and mid-infrared colors) stars with modest and moderate activity ($\mathrm{EW_{IRT}<0.5\;\mathrm{\AA}}$) appear to have colors similar to inactive stars. However, the most active stars with $J-K>0.6$ are significantly blueshifted in $N_\mathrm{UV}-V$ and redshifted in $W_1-W_2$. The bigger the offset from the main sequence the more active the star on average. 

A comparison between colors of active spectra and their first nearest inactive neighbors shows that stars above $\mathrm{EW_{IRT}>1\;\mathrm{\AA}}$ display a slight offset towards the red in the $J-K$ color ($0.07 \pm 0.10$) and a remarkable ultraviolet ($-0.93 \pm 1.019$ offset towards the blue in the $N_\mathrm{UV}-V$ color) and infrared excess ($0.037 \pm 0.079$ offset in the $W_1-W_2$ color). It is not entirely clear whether the red offset originates in the near infrared excess emission or reddening. A great fraction of active stars reside in the Galactic plane and some of them might still be embedded in their birth cocoons but in general distances to dwarfs are less than $400\;\mathrm{pc}$ due to the observational luminosity limit in RAVE. \citealp{2004ARA&A..42..685Z} state that for classical T Tauri stars (pre-main sequence stars less than a few million years old and associated with circumstellar nebulosity) ``Excess emission above photospheric emission at near-, mid- and far-infrared wavelengths suggests the presence of substantial quantities of nearby heated dust particles in the form of a disk or envelope or both. Ultraviolet line and continuum emission of classical T Tauri stars indicates that, typically, they are actively accreting a portion of the surrounding gas and dust." On the other hand, \citealp{2004ARA&A..42..685Z} say, ``by an age of $\sim 10 \;\mathrm{Myr}$, stellar optical activity is much reduced and near- and mid-infrared excesses are very infrequent". Dusty circumstellar disks (around pre-main sequence stars) and their infrared emission are discussed in detail by \citealp{2001ARA&A..39..549Z}.


According to \citealp{1999ARA&A..37..363F}, the X-ray emission is a ubiquitous characteristic of young low-mass stellar objects from protostars with $\sim10^5$ years to stars approaching the zero-age main sequence (ages of $10^7$ years) as a consequence of powerful magnetic reconnection flares. 
It is thus no surprise that very active red stars exhibiting X-ray flux 
overlap with the most active region of the diagrams. What is more, RAVE stars with $\mathrm{EW_{IRT}>1\;\mathrm{\AA}}$ identified in the Simbad database as pre-main sequence, T~Tauri and young objects, together with objects younger than $100\;\mathrm{Myr}$ (identification is given in the appendix, Table \ref{tab.reference_stars2}), reside in the same active part of the plots. Members of clusters, moving groups and associations are plotted for comparison.

The consistency of the photometric data justifies the assumption that the majority of the stars off the main sequence in the $N_\mathrm{UV}-V$ - $J-K$ and $W_1-W_2$ - $J-K$ planes (i.e. very active stars with $\mathrm{EW_{IRT}>1\;\AA}$) are very young, possibly still residing on the Hayashi tracks. The sample traces stars originating from recent star formation events in the Solar neighborhood. 
The reader may recall that active RAVE objects are field stars with a random selection function.
Since the active catalog consists of a great fraction of Tycho-2 stars in the Solar neighborhood it is expected that parallaxes provided by Gaia will enable their precise placement on the isochrones. Besides the study of activity on the pre-main sequence, the position on the isochrones will help to investigate the nature of solar-like active objects with $(J-K)<0.5$. They reach activities of up to $\sim 1 \;\mathrm{\AA}$ and do not photometrically differ from inactive stars (except for a mild redshift in $J-K$) despite their young age. They evolve faster since their mass is greater than the mass of late-type stars. Moreover, the nature of the generation of a magnetic dynamo might depend on stellar mass and the relative depth of the convective envelope.

\begin{figure}
\includegraphics[width=\columnwidth]{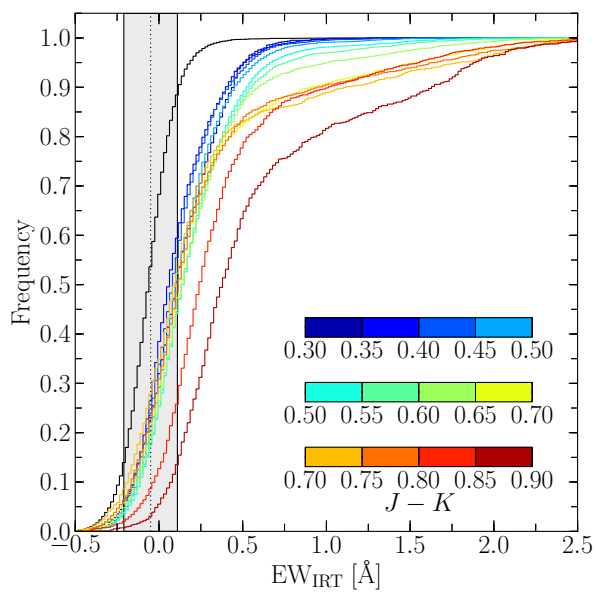} 
\caption{Cumulative distribution of activity for selected $J-K$ color ranges. The redder the population the higher the probability for activity and, at the same time, higher activity rates. The grey vertical band indicates the center and $\pm 1\sigma$ deviation of the activity distribution of inactive stars (depicted by the black curve).}
\label{fig.ewirt_distribution_cumulative_colors}
\end{figure}

\bigskip


\begin{figure}
\includegraphics[width=\columnwidth]{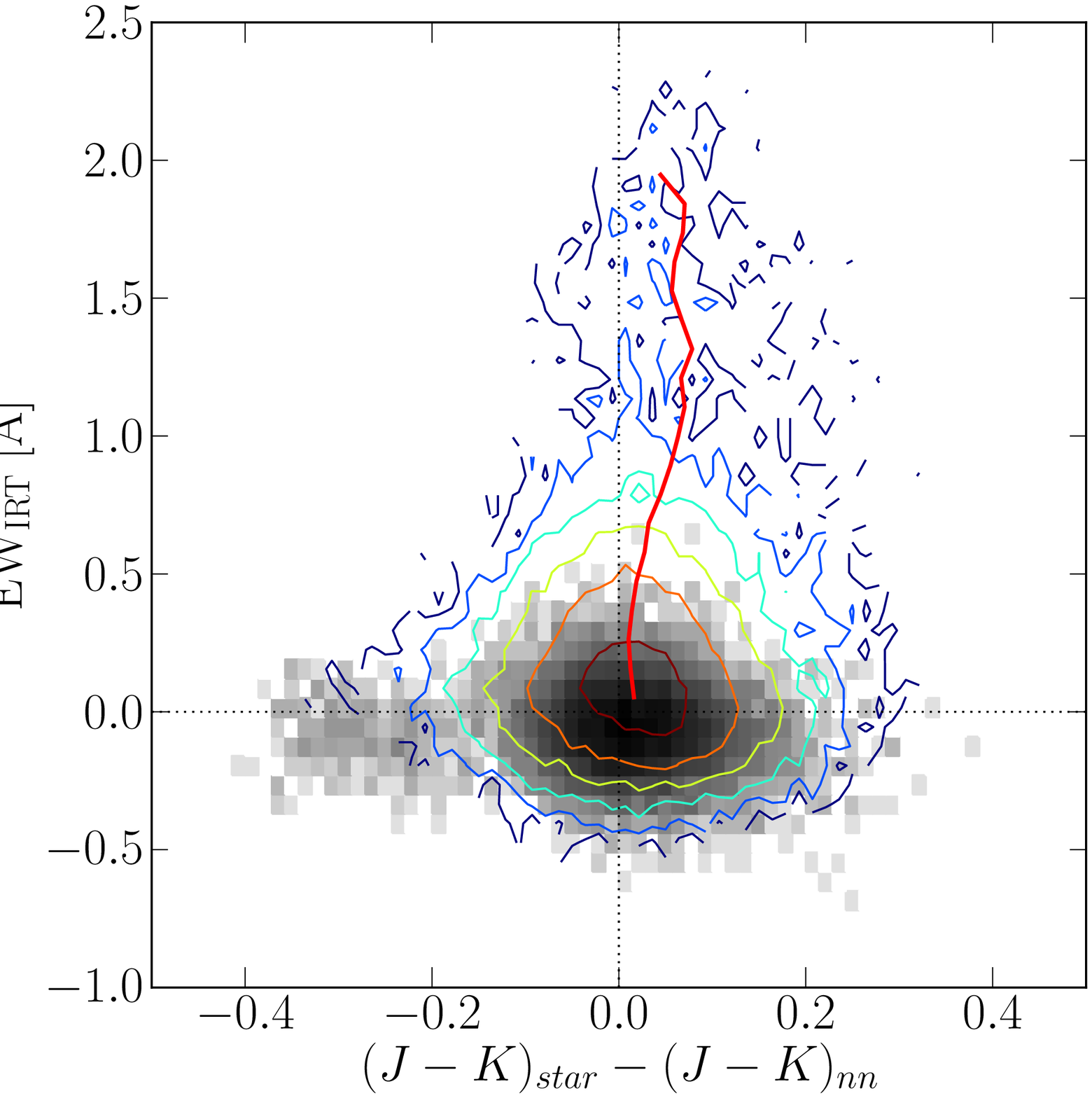} 
\caption{$\mathrm{EW_{IRT}}$ versus $J-K$ color excess for active stars (contours). The latter is given as a difference between the color of the star and the color of its nearest inactive neighbor assuming that reddening of inactive stars (grey histogram) is negligible. The red vertical solid line marks the average color excess at the given activity rate (the bin size is $0.11\;\mathrm{\AA}$). The majority of stars is concentrated around zero color excess. Very active stars are prone to mild reddening (or infrared excess emission) because they mostly reside in the Galactic plane and are possibly still embedded in their gas and dust cocoons (pre-main sequence stars).}
\label{fig.jmk_sosedje}
\end{figure}


Activity of cool dwarfs ($J-K\ge0.7$) increases on average with redder color (upper panel of Fig. \ref{fig.ewirt_jmk}: green line for active stars and blue for the inactive sample). Apart from the color-color diagrams, cumulative distributions over $\mathrm{EW_{IRT}}$ for subsets within selected color ranges (Fig. \ref{fig.ewirt_distribution_cumulative_colors}) confirm that the more active peak in the distribution consists mainly of red stars. The redder the population the more active the stars. Only the bluest stars have their distribution centred below $\mathrm{EW_{IRT}=0.1\;\AA}$ ($1\sigma$ above the center of the reference inactive distribution). The distribution spreads out toward higher activities above $J-K>0.7$. A significant fraction (20\%) of stars with $0.85 < J-K < 0.9$ is more active than $1\;\mathrm{\AA}$ as predicted earlier from the red sample.

According to the literature (e.g., R84) 
there exists a lower color-dependent boundary of the total surface flux in the Ca~II~H\&K lines of the main sequence stars that corresponds to the basal emission of the star. The basal activity remains constant for solar-like stars while emission values between $1<B-V<1.6$ are increased (\citealp{2010ApJ...725..875I}). The region with higher basal levels corresponds to $0.65<J-K<1$, which could explain the $\mathrm{EW_{IRT}}$--$J-K$ relation. 

However, basal emission is also expected to be present in the inactive sample as it is found in all stars with convective envelopes. This component should thus already have been eliminated from the $\mathrm{EW_{IRT}}$ by the spectral subtraction technique assuming that active and inactive spectra represent similar stars. Nonetheless, $\mathrm{EW_{IRT}}$ increases with color not only for active stars but for the inactive sample as well despite the fact that a star and its nearest inactive neighbor have almost identical colors (i.e., should have similar spectra) as shown in Figure \ref{fig.jmk_sosedje} (note that the nearest neighbor search was based solely on the normalized spectrum, other parameters (e.g. colors) were excluded). The activity of inactive red stars is on average lower than the emission levels of red active stars. The reason for the marginal activity detected in the red spectra of the inactive sample is due to relaxed selection criteria in the coolest part of the set. As shown in Figure \ref{fig.ewirt_distribution_cumulative_colors} the redder the star the more likely it appears to be active. At the same time, if nearest inactive neighbors in the $\mathrm{EW_{IRT}}$ determination of active stars exhibit mild excess emission, the $\mathrm{EW_{IRT}}$ of red active stars is underestimated. Mild activity of the inactive database is certainly one of the disadvantages of the spectral subtraction technique.
Nevertheless, it seems that the trend of increasing activity with color cannot be explained by the basal emission. Other underlying phenomena, perhaps connected to the nature of convection and the generation of magnetic fields in the coolest (and youngest) stars, must be present.

\section{Age -- activity relation} \label{sec.age}
The age of a star cannot be measured directly. It can only be estimated using various indirect (possibly distance-independent) methods. For pre-main sequence stars the lithium $6708\;\mathrm{\AA}$ depletion boundary is used (e.g. \citealp{2014EAS....65..289J}) while gyrochronology exploits the inverse correlation between the age and the rotation rate for main sequence F, G, K and early M stars (\citealp{2007ApJ...669.1167B}). Astroseismology proved very successful for pulsating stars but it is not entirely clear if the detection of pulsations in stars cooler than the Sun is possible (\citealp{2011ApJ...743..143H}). 
Due to the absence of sensitive indicators the evaluation of the main sequence field star ages remains fundamentally difficult.

Young solar-type stars appear to be more magnetically active than older, evolved stars. The magnetic activity level correlates with a diminishing stellar rotation rate through the coupling by the magnetic dynamo effect (\citealp{1972ApJ...171..565S, 1984ApJ...279..763N}, \citealp{1995ApJ...438..269B}). 
\citealp{1972ApJ...171..565S} 
studied cluster members and binary companions of more massive stars and reported the relationship between stellar age and chromospheric emission decay to be proportional to $\tau^{-1/2}$. Many authors later calibrated the power-law relationship with their samples and ages provided by other, alternative methods. While \citealp{2013MNRAS.431.2063S} confirm the relation for G-, K- and M-dwarfs, M08 review calibrations from the literature (\citealp{1987ApJ...315..264B, 1991ApJ...375..722S, 1999A&A...348..897L}) 
and describe their new empirical calibration for the Ca~II~H\&K, reliable for solar-type (F7-K2) stars between 0.6 and $4.5\;\mathrm{Gyr}$:

\begin{equation}
\log{\tau} = -38.053 - 17.912 \log{\mathrm{R'_{HK}}} - 1.6675 (\log{\mathrm{R'_{HK}}})^2.
\label{eq.mamajek}
\end{equation}

Here $\mathrm{R'_{HK}}$ is a ratio between the sum of emission fluxes in the Ca~II~H\&K lines (bandwidth of $1\;\mathrm{\AA}$) and the bolometric flux of the star.

The nature of the chromospheric activity evolution has been examined by various authors, e.g., \citealp{2013A&A...551L...8P, 2011AJ....141..107Z, 2005A&A...431..329L}. A comparison of the age-activity relation as a dating technique with other methods for young stars (lithium depletion boundary and gyrochronology) was described by \citealp{2014EAS....65..289J}. \citealp{2010ARA&A..48..581S} comprehensively reviews stellar dating techniques and provides advantages and disadvantages of using the activity -- age relation. He emphasizes the fact that the main parameter is stellar rotation. Its relation to activity is poorly understood as it turns out that especially for the pre-main sequence and zero-age main sequence stars a given rotation rate can result in a range of activity levels. 
\citealp{2013A&A...556A..36G} show the relation to converge only after $\sim 500\;\mathrm{Myr}$. An approximate tuning is possible for stars between $500\;\mathrm{Myr}$ and $1.5-2\;\mathrm{Gyr}$. Stars later remain active but it is more difficult to detect their diminishing low levels of activity.
A large spread in rotational velocities of the youngest stars disproportional to their ages escalates the complexity of the problem besides the evidence that high rotation rates with longer convective turnover times at lower masses and pre-main sequence stars cause saturation of chromospheric activity indicators.  
The activity lifetime of stars of the latest types (M3-M5) is increased, possibly because of an onset of the full convection and different dynamo mechanisms (\citealp{2008AJ....135..785W}). 
For solar-type stars older than $\sim 500\;\mathrm{Myr}$ accurately measured $\log{\mathrm{R'_{HK}}}$ values yield $\log{\tau}$ to $\pm 0.2\;\mathrm{dex}$ -- the precision is $\pm60\%$ (M08). The uncertainty grows for younger stars and dating becomes useless below $100\;\mathrm{Myr}$ where only upper age limits can be provided.

The influence of metallicity on the absorption line depths makes activity indices based on the ratio of Ca~II~H\&K flux and continuum (e.g., the S index, \citealp{1978PASP...90..267V} and $\log{\mathrm{R'_{HK}}}$, \citealp{1984ApJ...279..763N}) metallicity-dependent (\citealp{1998MNRAS.298..332R}). This effect is reduced in the RAVE active catalog with the spectral subtraction technique. However, lower abundances might affect the dynamo itself and the generation of the heating.


\subsection{$\mathrm{Age - EW_{IRT}}$ calibration} \label{sec.calibration}

\begin{figure*}
\begin{center}
\includegraphics[width=0.7\textwidth]{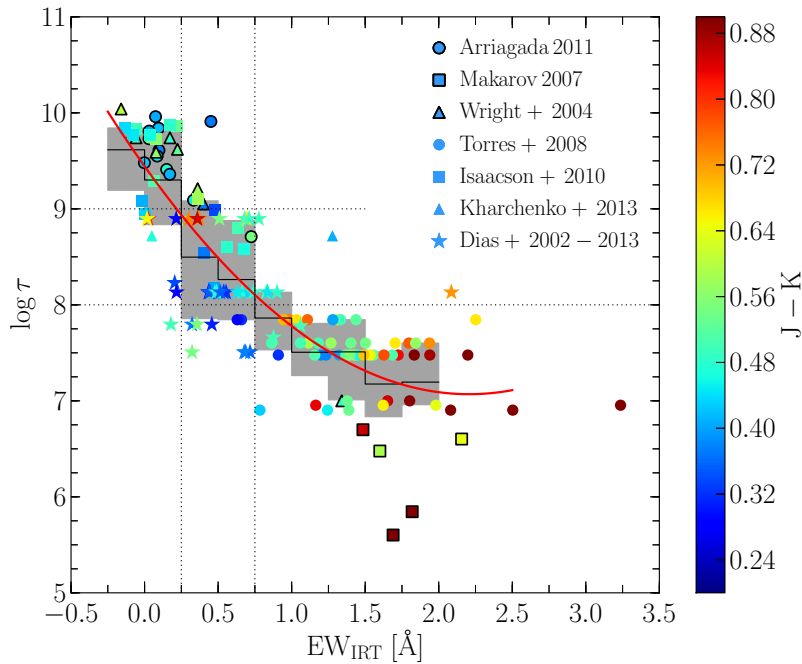} 
\caption{Calibration of the $\mathrm{age - EW_{IRT}}$ relation (red line). Ages (logarithmic scale, $\log{\tau}$) from external catalogs are shown against their $\mathrm{EW_{IRT}}$ values (dots). The colors indicate $J-K$ indices, the scale is given in the color bar on the right. A moving average of $\log{\tau}$ shows a steady decline of activity with age (black line; grey area is $1\sigma$ standard deviation). Dotted lines mark three main age--activity regimes: $\ge 1\;\mathrm{Gyr}$ ($\mathrm{EW_{IRT} <0.25 \; \AA}$), between 0.1 and $1\;\mathrm{Gyr}$ ($\mathrm{0.25\; \AA <EW_{IRT} <0.75 \; \AA}$) and younger than $100\;\mathrm{Myr}$ ($\mathrm{EW_{IRT} \ge 0.75 \; \AA}$).}
\label{fig.age_catalog}
\end{center}
\end{figure*}

\begin{table}
\begin{center}
\caption{Average age within the activity interval of the reference stars with known age (Fig. \ref{fig.age_catalog}).}
\begin{tabular}{r l c}
\hline
\hline
$\mathrm{EW_{IRT,1} \; [\AA]}$ & $\mathrm{EW_{IRT,2} \; [\AA]}$ & $\log{\tau}$ \\
\hline
-0.25 & 0.0 & $9.62^{+0.23}_{-0.42}$ \\
0.0 & 0.25 & $9.30^{+0.52}_{-0.47}$ \\
0.25 & 0.5 & $8.50^{+0.59}_{-0.64}$ \\
0.5 & 0.75 & $8.26^{+0.62}_{-0.42}$ \\
0.75 & 1.0 & $7.86^{+0.27}_{-0.31}$ \\
1.0 & 1.25 & $7.51^{+0.30}_{-0.24}$ \\
1.25 & 1.5 & $7.51^{+0.34}_{-0.50}$ \\
1.5 & 1.75 & $7.18^{+0.33}_{-0.34}$ \\
1.75 & 2.0 & $7.19^{+0.41}_{-0.24}$ \\
\hline
\end{tabular}
\label{tab.average_age_reference_stars}
\end{center}
\end{table}

Ages of RAVE stars have been estimated by \citealp{2014MNRAS.437..351B}. Because their age prior was focused mainly on older stars, active stars are estimated to be older than a few Gyr. In our work, reference ages of known Galactic open cluster members, moving groups and ages derived from the $\mathrm{Age - \log{R'_{HK}}}$ relation are used (see Table \ref{tab.reference_stars2} for references and additional info).

Despite many limitations described in the previous section and considerable scatter of the data the large scale correlation between activity and age is still evident in RAVE (Fig. \ref{fig.age_catalog}). It allows us to provide a coarse calibration. The goal of this work is not precise dating but a relation that gives an order of magnitude age estimate that can significantly contribute to the establishment of young candidate lists in large field star surveys.

A moving mean of the ages within activity bins ($0.25\;\mathrm{\AA}$) up to $2\;\mathrm{\AA}$ is presented together with the $\pm 1\sigma$ deviations in the Table \ref{tab.average_age_reference_stars}. Since this is an approximate relation we assume that more active stars are not older than a few tens of millions of years.

The ages of the oldest and the least active reference stars coming from the catalogs with known $\log{\mathrm{R'_{HK}}}$--age relation values reach a few billions of years. The Solar $\log{\mathrm{R'_{HK}}}$ activity is $\sim -4.9$ (M08) which translates to $-0.2 \mathrm{\AA} < \mathrm{EW_{IRT}} < 0.2 \mathrm{\AA}$ according to the $\log{\mathrm{R'_{HK}}} - \mathrm{EW_{IRT}}$ correlation in Fig. 7 in Z13. The age of a star within this activity bin is between 1 and $6 \; \mathrm{Gyr}$.

Regarding the discussion in Sec. \ref{sec.photometry} spectra with activity above $\sim 1 \;\mathrm{\AA}$ and $J-K \gtrsim 0.5$ probably belong to the pre-main sequence population of stars and they must thus be younger than $\sim 100\;\mathrm{Myr}$. The more active peak in the activity distribution occurs around $1.5\;\mathrm{\AA}$ which translates to a few tens of millions of years. 
Although emission levels below $\sim 1 \;\mathrm{\AA}$ on average correspond to ages older than $100\;\mathrm{Myr}$, the distribution of activity of young stars with $J-K<0.5$ extends down to $\sim -0.5 \; \mathrm{\AA}$. The relation is color-dependent as pointed out by many authors, e.g., M08. Nevertheless, the correlation between age and activity in Figure \ref{fig.age_catalog} does not show any particular color trend. The reason is probably a combination of the fact that most of the calibration stars are solar-like with $J-K < 0.5$ and that the scatter is too large.


\bigskip

A parabola was used to find the age of a star for a given activity level, as follows:
\begin{equation}
\log{\tau} = 0.49 \, \mathrm{\AA^{-2}} \, \mathrm{EW_{IRT}^2} -2.15 \, \mathrm{\AA^{-1}} \, \mathrm{EW_{IRT}} + 9.44.
\label{eq.age_parabolic_fit}
\end{equation}

The fit is only informative. Clearly, the uncertainties of the derived ages from the equation are large and this method is not to be used for precise dating. 
Due to the large scatter of the data used in this paper only a categorization of ages into three orders of magnitude is possible: Ages of stars with $\mathrm{EW_{IRT} <0.25 \; \AA}$ are estimated to be 1 billion years or greater. Stars with activity levels between 0.25 and $0.75\; \mathrm{\AA}$ are estimated to be between 0.1 and $1\;\mathrm{Gyr}$ old whereas the most active stars ($\mathrm{EW_{IRT} \ge 0.75\; \AA}$) are younger than $100\;\mathrm{Myr}$. Nevertheless, the result is highly beneficial in large surveys for the identification of a large number of young stars that are candidates for further more detailed investigation using alternative methods.




\bigskip
The catalog of active stars will appear online in the Vizier database. The description of the columns including the age estimate is given in Table \ref{tab.data}.

\begin{table}
\centering
\caption{Description of the columns in the data table.}
\begin{tabular}{l | l | l}
\hline
\hline
Column & Units & Description \\
\hline
RA & deg & Right ascension (J2000.0) \\
DE & deg & Declination (J2000.0) \\
S/N & & Signal-to-noise ratio of spectrum \\
EWirt & $\mathrm{\AA}$ & $\mathrm{EW_{IRT}}$ \\
eEWirt & $\mathrm{\AA}$ & Uncertainty of the $\mathrm{EW_{IRT}}$ \\
LogAge & & $\log{\tau}$ estimate from the age--activity relation \\
\hline
\end{tabular}
\label{tab.data}
\end{table}

\section{RAVE distances} \label{sec.distance}
\begin{figure*}
\includegraphics[width=\linewidth]{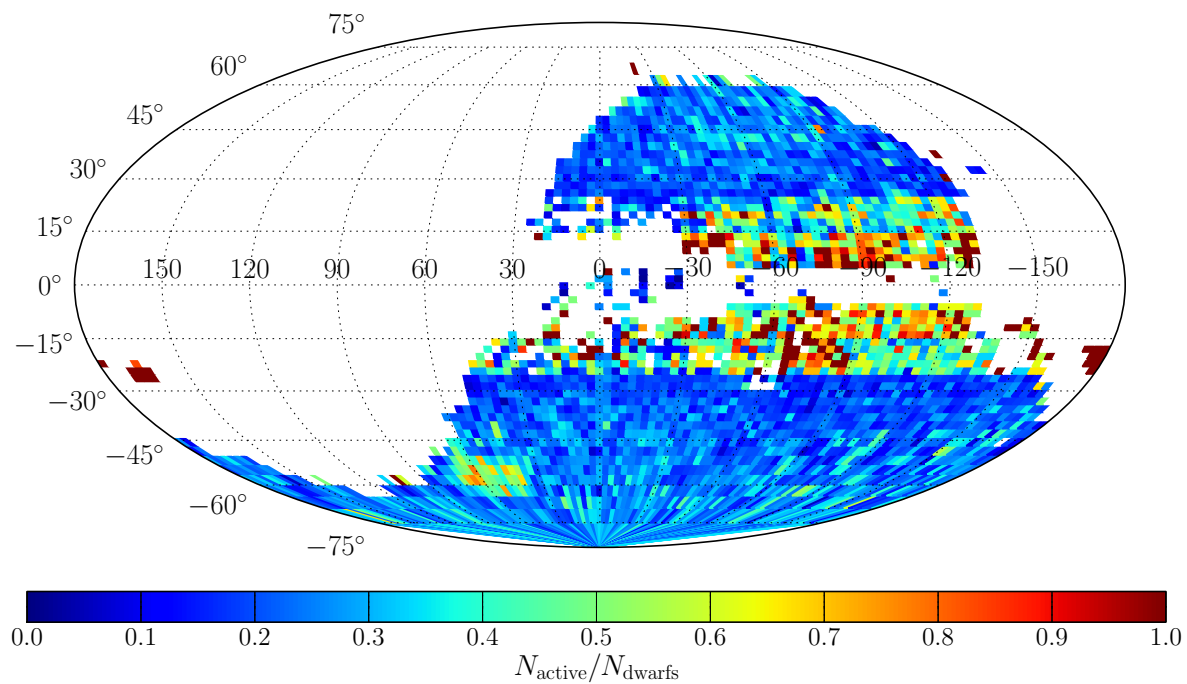} 
\caption{The number of all active candidates over the number of all RAVE dwarfs with $\log{g}>3.75$. The occurrence of activity is high in the Galactic plane, partly because of the color-cut $J-K>0.5$ performed at $25\degree<|b|$. An overdensity of active stars in the direction of the Aquarius constellation ($35\degree<l<75\degree$ and $-62\degree<b<-50\degree$) is seen due to additional color-limited observations of the Aquarius stream.}
\label{fig.aquarius_galaxy_overdensity}
\end{figure*}

\begin{figure*}
\begin{center}
\includegraphics[width=0.8\textwidth]{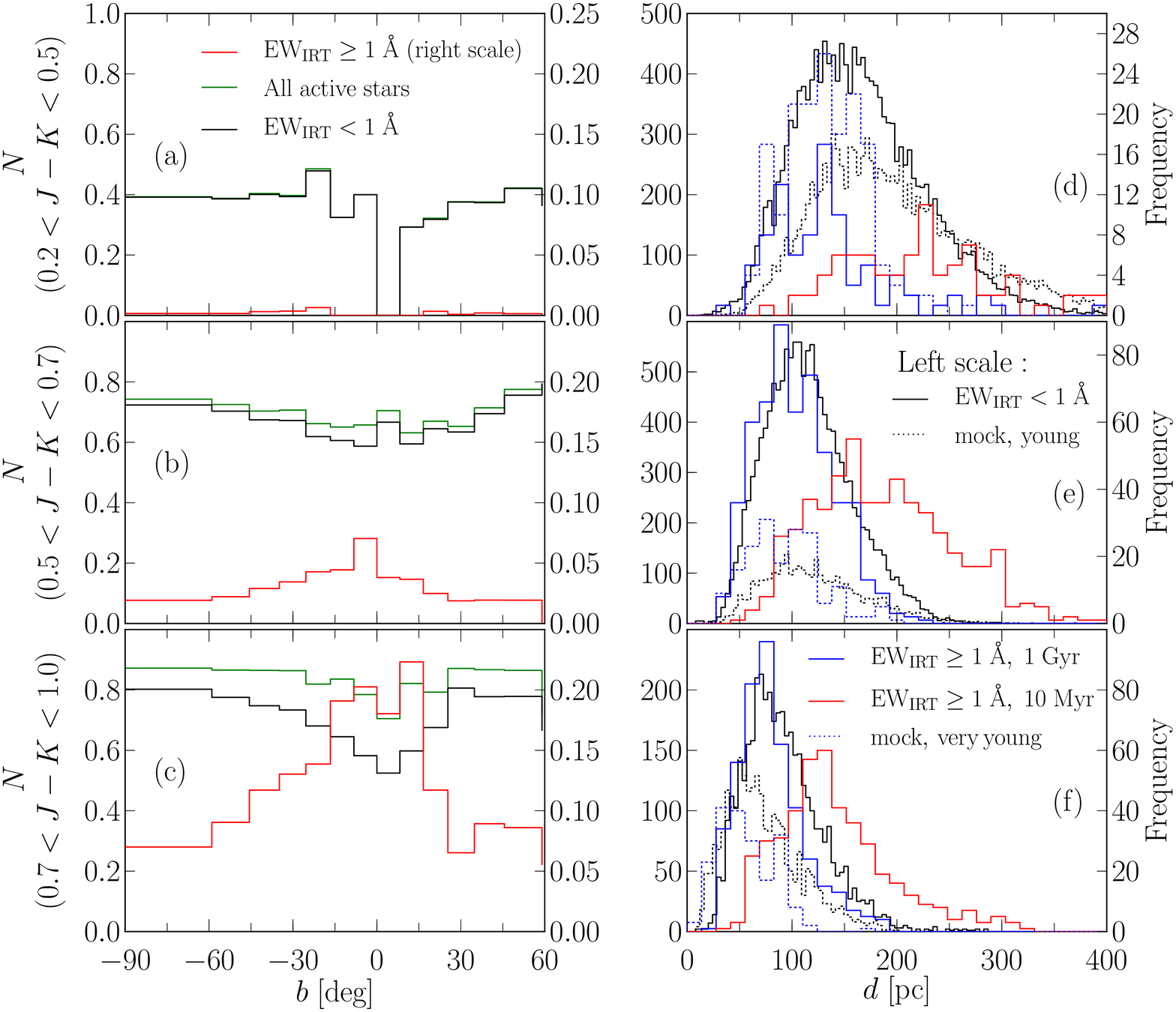} 
\caption{\textit{Left}: The number of active stars over the number of all RAVE dwarfs with $\log{g}>3.75$ versus the Galactic latitude for 3 color classes. Uneven bin sizes correspond to areas with equal surface in the sky. The most active stars (red line; scale on the right) concentrate at low Galactic latitudes. Distributions for all active stars (green line) and stars with $\mathrm{EW_{IRT} < 1\;\AA}$ (black line) are given for reference (their scale is on the left). \textit{Right}: Isochronal \textit{heliocentric} distance estimates at 1 Gyr. Because more active stars (especially solar-like ones with $0.2<J-K<0.5$, panel d) on average have smaller distances than less active stars, an additional $10\;\mathrm{Myr}$ distance was determined as an upper limit for this group as it is evident that they might still reside in their pre-main sequence phases. Histograms for stars with $\mathrm{EW_{IRT} < 1\;\AA}$ (solid and dotted black line) are shown according to the left axis while the scale for stars with $\mathrm{EW_{IRT} \ge 1\;\AA}$ (solid and dotted blue and red line) is shown on the right. Distribution of mock-RAVE stars (using the Galaxia model) is shown for comparison (dashed lines; `young stars' mark sample of objects with ages corresponding to $\mathrm{0.1\;\AA < EW_{IRT} < 1\;\AA}$, `very young stars' would show $\mathrm{EW_{IRT} > 1\;\AA}$ at their age.}
\label{fig.b_d_z}
\end{center}
\end{figure*}

The number of all active stars over the number of all RAVE dwarfs with $\log{g}>3.75$ (Figure \ref{fig.aquarius_galaxy_overdensity}) indicates that star formation mainly occurs in the Galactic plane, as expected. Additionally, a plot of the ratio between the number of active stars over the number of all observed RAVE dwarfs for three color-classes versus the Galactic latitude (Fig. \ref{fig.b_d_z}, panels a, b and c) demonstrates that the most active stars (red line) are most probably found closer to (and in) the Galactic plane. However, because the stars at low latitudes can theoretically still lie at great distances from the Galactic plane $z$, it is more convenient to plot the distribution over $z$.

Spectrophotometric distances to the RAVE stars have been estimated by an updated version of the \citealp{2014MNRAS.437..351B} distance pipeline which includes a wider range of metallicities (the lower limit was extended from [Fe/H]=-0.9 in the previous version down to -2.2~dex) in the adopted isochrones.
A comparison of distances from RAVE and external catalogs (for cluster and moving groups members) shows an offset (the average difference between distances for a particular star closer than $500\;\mathrm{pc}$ is $31\;\mathrm{pc}$) and a rather big scatter (standard deviation of the difference between the estimates is $98\;\mathrm{pc}$ and $36\;\mathrm{pc}$ for stars closer than $100\;\mathrm{pc}$). A color-magnitude diagram using RAVE distances reveals that a large fraction of very active stars (above $J-K>0.6$) reside either above giants (their surface gravity is typically underestimated) or below the main sequence.

To overcome the overall problem of distances, stars were confined between a $1\;\mathrm{Gyr}$ isochrone on the main sequence and a $10\;\mathrm{Myr}$ isochrone in the pre-main sequence zone (it corresponds to $\sim 1.5 \;\mathrm{M_\odot}$ main-sequence turn-on initial mass; more massive stars have already settled on the ZAMS while stars below this point are still contracting). The latter age is likely underestimated but it serves as the lower limit. Theoretically, for the reddest stars the two distances differ for a factor $\sim 1.7$ and for $\sim 1.9$ for the bluest stars. Both isochrones use Padova models (PARSEC release v1.2S; \citealp{2012MNRAS.427..127B}); the metallicity is set to the Solar value. 

The detection limit for K4-type RAVE dwarfs is around $200\;\mathrm{pc}$ ($\sim 400\;\mathrm{pc}$ for solar-like dwarfs).
The vertical scale height of the Galactic thin disk is $\sim 300\;\mathrm{pc}$ (\citealp{2002ARA&A..40..487F}) which means that practically all active candidates are expected to be situated within this layer. Due to the $I$ magnitude being the only limit outside the color-selected Galactic plane, populations of RAVE stars with similar spectroscopic types should in principle occupy volumes in space of the similar size. For this reason the data analysis was performed for three different color ranges. Moreover, active stars were divided into two groups: less active stars with $\mathrm{EW_{IRT}<1\;\AA}$ and more active stars with $\mathrm{EW_{IRT} \ge 1 \AA}$. Distances for inactive stars are not considered because of the selection criterion: very high signal-to-noise values of template spectra were demanded when building the inactive database which effectively means that these stars must reside very close to the Sun.

At $1\;\mathrm{Gyr}$ (Fig. \ref{fig.b_d_z}, panels d, e and f) more active stars (especially solar-like ones with $0.2<J-K<0.5$, panel d) on average have smaller distances than less active stars. There is no reason for their absence at greater distances. As it is evident that the more active stars might still reside in their pre-main sequence phases, the problem can be solved by raising the most active subset to younger and more luminous isochrones as stellar distance in the Hertzsprung-Russell diagram grows in the direction toward the giants. The red line in the plot represents the distance distribution for more active stars at $10\;\mathrm{Myr}$. While distances for both activity groups with $J-K>0.5$ (panels e and f) at $1\;\mathrm{Gyr}$ are comparable, distances for more active stars at $10\;\mathrm{Myr}$ are expectedly larger. However, it seems that the distance distribution for very active solar-like stars ($0.2<J-K<0.5$, panel d) is in disagreement with their placement on the color-color diagram (Fig. \ref{fig.photometry_young}). Photometry shows that this group of stars is placed on the main sequence (so they must be put on the $1\;\mathrm{Gyr}$ isochrone). One might think that the possible reason for the lack of more distant stars could be the detection limit of activity: because solar-like stars show lower levels of emission its detection is more challenging in noisy spectra, i.e., spectra of more distant stars. Nevertheless, this is not the case since the distribution of less active stars reaches greater distances.

In order to better understand the distance distribution of the youngest RAVE stars, the population synthesis code Galaxia (\citealp{2011ApJ...730....3S}) was used to produce a mock-RAVE catalog (\citealp{wojno_2016}). First, a full-sky synthetic catalog was generated with an apparent magnitude limit $0 <
I_\mathrm{DENIS} < 14$. These apparent magnitudes were corrected for extinction using Schlegel map values (\citealp{1998ApJ...500..525S}) at the position of the star, and modified to include errors using a RAVE-like magnitude error distribution. The RAVE selection function ($S_{\mathrm{select}} \sim (l,\, b,\, 
I_{\mathrm{2MASS}},\, J-K_s)$) was then applied to this sample to match the distribution of the observed RAVE stars. Because it is evident that active stars are rare in the blue part of RAVE's main sequence, the upper $\log{\mathrm{g}}$ limit was set to 4.4 and $T_\mathrm{eff}<6000\;\mathrm{K}$. All stars on the lower main sequence satisfying these criteria are included in the mock sample because activity is not simulated in the catalog. Further, only the youngest stars were selected from the set: stars below $\log{\mathrm{age}}=7.78$ (purple shade in the plot; this age limit corresponds to $\mathrm{EW_{IRT} > 1\;\AA}$ according to Eq. \ref{eq.age_parabolic_fit}) and stars with $7.78<\log{\mathrm{age}}<9.23$ ($\mathrm{0.1\;\AA < EW_{IRT} < 1\;\AA}$; green line). In order to additionally divide the sample into three color classes the temperature was converted to $J-K$ colors (stars with $T_\mathrm{eff}>5100\;\mathrm{K}$ correspond to $J-K<0.5$, stars with $4500\;\mathrm{K}<T_\mathrm{eff}<5100\;\mathrm{K}$ to $0.5<J-K<0.7$ and stars cooler than 4500~K to $J-K>0.7$).

A qualitative comparison between the observed and mock catalog distributions shows similar trends. The lack of more distant, very young stars in the bluest sample ($0.2<J-K<0.5$, panel d) is also evident in the mock sample. The nonexistence of distant, very young stars in the mock catalog as well excludes the selection effect in the observed RAVE sample.

Note that there is a quantitative difference present between the mock and observed dataset. A possible reason is that Galaxia does not inherently include errors on distances, and that isochronal estimates in this work do not take extinction into account.

Despite the efforts to estimate stellar distances their values remain unreliable and inconclusive. For this reason we skip the study of the distribution of stars over the distance from the Galactic plane $z$. Accurate and precise parallaxes from the Gaia catalogs will provide more detailed insight into the distribution of active stars over the distance from the Galactic plane\footnote{\textit{Note to the reader:} this paper was submitted before the TGAS catalog (Gaia-DR1) was released.}.


\section{Discussion and conclusions} \label{sec.discussion}
This work reports on the improved identification of stellar activity in the catalog introduced by Z13 and extends its quality control. Over 9,000 unreliable or improper spectra were removed from the sample (Sec. \ref{sec.data}). 2919 new spectra of 2882 stars published for the first time in DR5 were found to show signs of activity and were added to the active catalog. The increasing excess emission flux in the Ca~II~IRT in the averaged spectra for each activity bin from levels below the detection limit to individual cases with emission filling in calcium lines confirms the reliability of the data (Fig. \ref{fig.average_spectrum}).
The new catalog comprises of 38,678 candidate spectra with activities of almost 13,000 stars above $2\,\sigma$ and almost 22,000 stars above the $1\,\sigma$ emission detection level.

\bigskip
The distribution of activity is bimodal (Fig. \ref{fig.ewirt_distribution}). This statement is supported by the comparison of the active sample and a single Gaussian function equivalent to the distribution of inactive stars. Since the number of stars around the peak at $\sim 1.5 \;\mathrm{\AA}$ exceeds the expected number by a few thousand times (Table \ref{tab.ewirt_above_sigma}), this is clear evidence for the existence of an additional peak in the distribution. Bimodality was also known to appear in the distribution of emission in the Ca~II~H\&K lines but the reason for the two peaks is not entirely clear. Since \citealp{2006AJ....132..161G} report a disappearing bimodality for stars with $[M/H]<-0.2$ different dynamo-generating mechanisms combined with metallicity dependence and stellar evolution might play a role in the structure of stellar magnetic fields and the resulting excess emission flux. A promising upcoming detailed chemical abundance information of up to 30 elements for $\sim 1$~million stars from the Galah Survey with a precision of $<0.1\;\mathrm{dex}$ (\citealp{2015MNRAS.449.2604D}) and with a combination of H$\alpha$ and H$\beta$ emission looks suitable to help to disentangle the underlying processes.

\bigskip

An interesting phenomenon in the distribution of active stars in the $\mathrm{EW_{IRT}}$ versus $J-K$ space is observed (Sec. \ref{sec.photometry}). Above $J-K>0.7$, the average activity is not only the highest but correlated with color as well (Fig. \ref{fig.ewirt_jmk}). The contribution of the color-dependent basal emission is excluded from the $\mathrm{EW_{IRT}}$ by the spectral subtraction technique using a measured inactive template database. A similar trend is observed in the sample of inactive stars (a small fraction of the coolest marginally active stars was recognized as inactive because of relaxed selection criteria due to the lack of inactive red stars). Further photometric investigation (Fig. \ref{fig.photometry_young}) unveiled their position off the main sequence in both the $J-K$ -- $N_\mathrm{UV}-V$ and $J-K$ -- $W_1-W_2$ space with exposing a large near-ultraviolet and mid-infrared excess emission. Pre-main sequence, T~Tauri and young stars together with X-ray sources (additional data from the Simbad database) and stars younger than $100\;\mathrm{Myr}$ (see appendix for the age reference) overlap with these regions. Undoubtedly, red and very active stars appear to be extremely young, probably still on their way to the main sequence. Again, the correlation between the activity and color of the coolest stars questions the nature of magnetic fields affected by stellar mass and the structure of convective envelopes.





\begin{figure}
\includegraphics[width=\columnwidth]{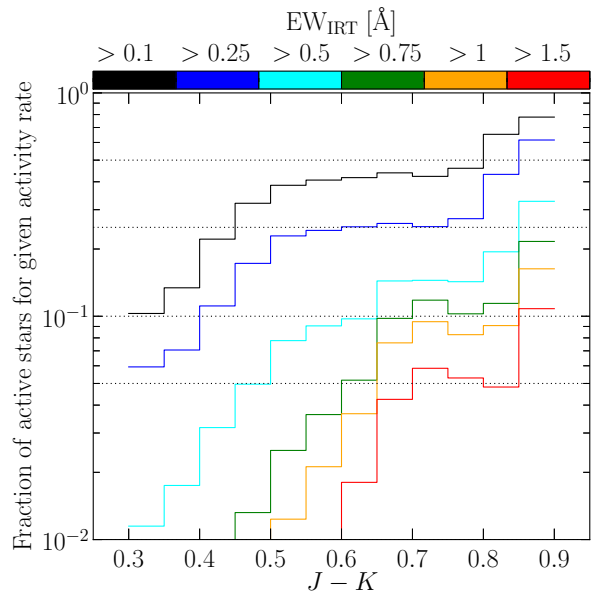} 
\caption{Probability for any detectable signs of activity present in the RAVE's Ca~II~IRT above a selected $\mathrm{EW_{IRT}}$ treshold for a randomly observed dwarf with the known $J-K$ in the Solar neighborhood within $\sim 300\;\mathrm{pc}$. The ratio of number of active stars above selected activity rate over the number of all RAVE dwarf stars is shown for each color bin. Dotted horizontal lines mark 5, 10, 25 and 50\% levels.}
\label{fig.ewirt_distribution_ratio}
\end{figure}

\bigskip
In order to further investigate the young age of the sample, a study of the age--$\mathrm{EW_{IRT}}$ relation and its calibration is presented as the main result of this paper (Sec. \ref{sec.age}). A clear correlation between external catalog ages and $\mathrm{EW_{IRT}}$ (Fig. \ref{fig.age_catalog}) enabled a coarse calibration of the relation on a large scale (Eq. \ref{eq.age_parabolic_fit}) because on smaller scales the scatter is significant. We introduce 3 age classes (above $1\;\mathrm{Gyr}$, between $1\;\mathrm{Gyr}$ and $100\;\mathrm{Myr}$ and younger than $100\;\mathrm{Myr}$) based on the stellar emission levels. 
Despite the large uncertainties the $\mathrm{EW_{IRT}}$ obtained from a single spectroscopic observation of a mid-range resolving power enables the identification of extremely young stars. The latter are of a special interest for the establishment of candidate catalogs in the large surveys and further detailed study due to the possibility to host exoplanets in their young stages besides recent star formation history.
Figure \ref{fig.ewirt_distribution_ratio} presents the probability for the activity rate for a randomly observed dwarf in the Solar neighborhood with known $J-K$ color. At $J-K>0.8$ more than 60\% of the stars show signs of activity detectable in RAVE and almost 10\% of the stars have $\mathrm{EW_{IRT}>1\;\AA}$ (i.e., are younger than $100\;\mathrm{Myr}$).

Solar-like stars settle on the main sequence sooner and evolve faster than very red dwarfs which makes the age--activity relation color-dependent. The population of the most active RAVE stars confirms the theory as solar-like stars with $\mathrm{EW_{IRT}>0.75\;\AA}$ younger than $100\;\mathrm{Myr}$ already reached the main sequence (Fig. \ref{fig.photometry_young}). The color dependence of the relation is not taken into account in this work because no trend is evident in the data.


\begin{table}
\centering
\caption{Color-dependent distribution of the youngest RAVE stars over the age classes.}
\begin{tabular}{l | l l}
\hline
\hline
 & $\mathrm{Age < 1 \; Gyr}$ & $\mathrm{Age < 100 \; Myr}$ \\
$J-K$ & $N \;(N/N_\mathrm{RAVE})$ & $N \;(N/N_\mathrm{RAVE})$ \\
\hline
$0.2-0.5$ & 5274 (0.055) & 216 (0.002) \\
$0.5-0.7$ & 6135 (0.233) & 811 (0.031) \\
$0.7-0.9$ & 3203 (0.431) & 932 (0.125) \\
\hline
Total & 14,612 (0.113) & 1959 (0.016) \\
\hline
\end{tabular}
\raggedright{\newline Note: A ratio of the number of active stars ($N$) over the number of all RAVE dwarfs ($N_\mathrm{RAVE}$) in a particular bin (130,000 altogether) is presented in the brackets. Note that the total amount of active stars (the bottom line) is slightly larger than the sum of the color-dependent numbers because not all stars have the 2MASS photometry available.}
\label{tab.age_distribution}
\end{table}




%

\bigskip
The distribution of ages for all candidate active RAVE stars is shown in Table \ref{tab.age_distribution}. 
$\sim$14,000 stars (11\% of all RAVE dwarfs) are estimated to be younger than $1\;\mathrm{Gyr}$ and $\sim$2000 (1.6\%) younger than $100\;\mathrm{Myr}$. The majority of the youngest stars are very red ($J-K>0.5$) and probably lies off the main sequence while stars younger than $1\;\mathrm{Gyr}$ tend to be hotter, as expected from Figures \ref{fig.photometry_young} and \ref{fig.ewirt_distribution_ratio}. The fraction of the coolest stars younger than $100\;\mathrm{Myr}$ (0.125) in the RAVE sample is $\sim$three times higher than the fraction in the mock-RAVE catalog (0.038).
The sample is a tracer of (a) recent large-scale star formation event(s) that occurred in the Solar neighborhood in the southern Galactic hemisphere, possibly due to the crossing(s) of the spiral arm(s).
\citealp{2004ARA&A..42..685Z} (and references therein) claim that a connection between nearby, young stellar groups and Scorpius-Centauri region (the nearest site of massive star formation) is evident and that, ``unless the Scorpius-Centauri mass function is very different from those of other star-forming regions, there must have been a handful of supernova explosions that triggered and later dispersed the maternal clouds of the young nearby groups".
An analysis with Gaussian mixture models done by \citealp{2009ApJ...700.1794B} on Galactocentric velocities of nearby Hipparcos stars (with distances less than $100\;\mathrm{pc}$) revealed that a large fraction of nearby stars (about 40\%) live their lives in a small number of moving groups. 
\citealp{2013A&A...552A..27M} analysed a sample of $\sim 1000$ stars from several catalogs of active stars (both Ca~II~H\&K and ultraviolet data) that have kinematic data available. They showed that areas with higher density and activity higher than average in the $U-V$ velocity plane correlate with the known positions of moving groups and clusters.
\citealp{2012MNRAS.426L...1A} detected significant overdensities in the velocity distributions of the RAVE disc stars in and beyond the Solar neighborhood (they report on the main local kinematic groups to be large-scale features surviving at least up to $\sim 1\;\mathrm{kpc}$ from the Sun).
According to their findings, a large number of young RAVE stars probably belong to known (or possibly still undiscovered) moving groups and recently disrupted clusters.
Orbital simulations would provide more detailed insight into this matter. Their photometric distances remain unreliable (Sec. \ref{sec.distance}), but around 50\% of the sample is found in the TGAS catalog. 
Since the database consists of dwarfs in the Solar neighborhood their parallaxes and proper motions are expected to be derived with high accuracy and precision. Distances will enable the accurate placement of objects on the isochrones. The RAVE sample of young active stars thus represents a valuable dataset to study not only the recent star formation history but the stellar physics of young and cool objects as well. 

\section*{Acknowledgements}
Funding for RAVE has been provided by: the Anglo-Australian Observatory; the Leibniz-Institut fuer Astrophysik Potsdam; the Australian National University; the Australian Research Council; the French National Research Agency; the German Research foundation; the Istituto Nazionale di Astrofisica at Padova; The Johns Hopkins University; the National Science Foundation of the USA (AST-0908326); the W.M. Keck foundation; the Macquarie University; the Netherlands Research School for Astronomy; the Natural Sciences and Engineering Research Council of Canada; the Slovenian Research Agency; Center of Excellence Space.si; the Swiss National Science Foundation; the Science \& Technology Facilities Council of the UK; Opticon; Strasbourg Observatory; and the Universities of Groningen, Heidelberg and Sydney. The RAVE web site is at \href{http://www.rave-survey.org}{http://www.rave-survey.org}.

This research made use of the cross-match service, the SIMBAD database and the VizieR catalog access tool provided by CDS, Strasbourg.

This publication makes use of data products from the Two Micron All Sky Survey, which is a joint project of the University of Massachusetts and the Infrared Processing and Analysis Center/California Institute of Technology, funded by the National Aeronautics and Space Administration and the National Science Foundation.

\bigskip

\bibliographystyle{apj}
\bibliography{bibliography}

\pagebreak

\appendix

\section{Reference stellar ages}
One hundred and thirty seven active RAVE spectra were used for the calibration of the age--activity relation. A small portion of spectra are repeated measurements of the same object. Such cases are treated independently in the age--activity relation calibration procedure.
Table \ref{tab.reference_stars2} lists all active RAVE candidates where external ages are available in the literature.

\begin{figure*}
\begin{center}
\includegraphics[width=\textwidth]{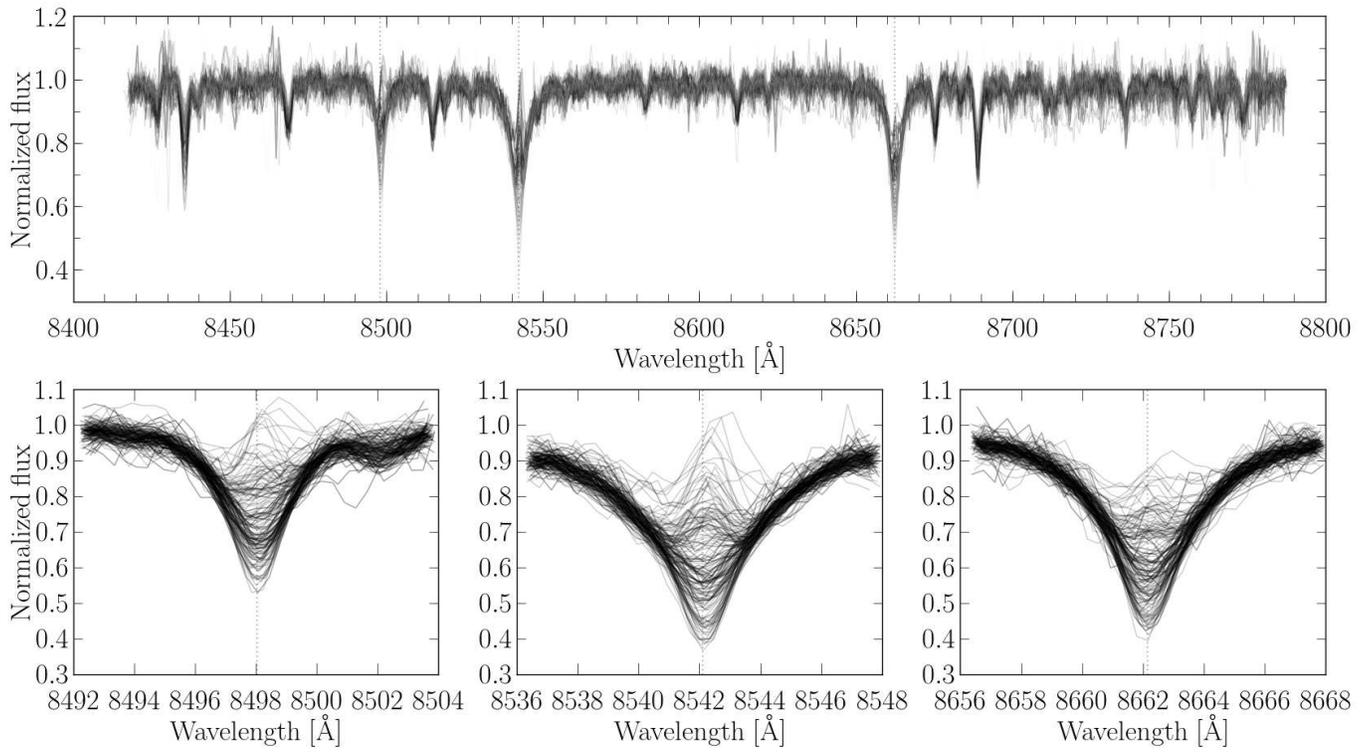}
\caption{Normalized RAVE fluxes of 137 active spectra used in the age--activity calibration. Their parameters together with age are listed in the Table \ref{tab.reference_stars2}. Calcium lines are shown separately in bottom panels.}
\label{fig.spectra_with_ages}
\end{center}
\end{figure*}

All spectra are depicted in Figure \ref{fig.spectra_with_ages} and show adequate behaviour, i.e. no peculiarities other than emission in calcium. Their activity levels range from negative values up to $\sim\mathrm{3\;\AA}$.

The first two columns of the table list right ascension and declination of a star (J2000.0), followed by the signal-to-noise (STN per pixel) estimation. Only spectra with STN$>20$ are included. 
Equivalent widths of excessive flux $\mathrm{EW_{IRT}}$ are a quantitative evaluation of emission levels. 
The references for cluster or moving group membership are the same as for the age if not stated otherwise. The references are marked as follows: A (\citealp{2011ApJ...734...70A}), C (\citealp{2014A&A...562A..54C}), D (\citealp{2012yCat....102022D}), I (\citealp{2010ApJ...725..875I}), K (\citealp{2013A&A...558A..53K}), M (\citealp{2007ApJ...658..480M}), T (\citealp{2008hsf2.book..757T}) and W (\citealp{2004ApJS..152..261W}).
Where $\log{\mathrm{R'_{HK}}}$ is known, ages in the original papers were derived from their $\log{\mathrm{R'_{HK}}}$--age relations. 
Ages of 7 stars were derived by 2 different teams. We list both values in such cases. Ages for 1 such star (Hyades member) are estimated to be of different orders of magnitude (and differ from the generally accepted Hyades age) while ages for the rest of them are comparable.
Other values rely on the age estimated from known stellar clusters and moving groups (column 'Group'; the membership reference is the same as for the age, if not stated otherwise). Two stars were found to be candidate members of two moving groups (Argus and IC~2391) at the same time. 
The table is supplemented by $N_\mathrm{UV}-V$, $B-V$, $J-K$ and $W_1-W_2$ colors. For a discussion on this matter see Sec. \ref{sec.photometry}.
Additional information is provided by the columns 'Sp. Type' and 'Simbad Type' marking spectral types and morphological classification available on the online Simbad database. Abbreviations for Simbad types are: PMS (Pre-main sequence star), X (X-Ray source), UV (UV-emission source), IR (Infra-Red source), hPM (High proper-motion Star), Cl (Star in Cluster), Var (Variable Star), Fl (Flare Star), B (Star in double system), Multi (Double or multiple star), Em (Emission-line Star), Y (Young Stellar Object) and sub-mm (sub-millimetric source).



\pagebreak
\clearpage
\begin{landscape}
\LongTables
\begin{center}
\begin{longtable}{l l | l | l | l l | l l l l | l l}
\caption{Table of 137 stars with known ages from the literature} \\
\hline
\hline
RA & DE & S/N & $\mathrm{EW_{IRT}}$  & Age & Group/$\log{\mathrm{R'_{HK}}}$ & $N_\mathrm{UV}-V$ & $B-V$ & $J-K$ & $W_1-W_2$ & Sp. Type & Simbad Type \\ 
$\mathrm{[deg]}$ & $\mathrm{[deg]}$ &  & $\mathrm{[\AA]}$ & $\mathrm{[Myr]}$ & & $\mathrm{[mag]}$ & $\mathrm{[mag]}$ & $\mathrm{[mag]}$ & $\mathrm{[mag]}$ &  &  \\
\hline
\endhead
240.84862 & -17.86169 & 29 & 1.69 & $\mathrm{0.4^M}$ & Sco Cen Upper &  & 1.58 & 1.02 & 0.090 &  M2 &  PMS, X, IR \\
239.75871 & -18.73725 & 35 & 1.82 & $\mathrm{0.7^M}$ & Sco Cen Upper &  & 1.28 & 0.90 & 0.057 &  K7 &  PMS, T Tau, X, IR \\
235.27850 & -26.94056 & 64 & 1.60 & $\mathrm{3^M}$ & Sco Cen Upper &  & 0.89 & 0.59 & -0.021 &  K1V &  PMS, X, IR \\
240.16913 & -22.00886 & 44 & 2.16 & $\mathrm{4^M}$ & Sco Cen Upper &  & 0.98 & 0.64 & -0.019 &  G9 &  X, PMS, T Tau, IR \\
226.48692 & -43.20089 & 28 & 1.48 & $\mathrm{5^M}$ & Lupus &  &  & 0.85 & -0.002 &  K7 &  Y, T Tau, IR, X, Star \\
62.98200 & -58.02986 & 35 & 0.78 & $\mathrm{8^T}$ & Octans & 4.7 & 0.19 & 0.43 & -0.037 &  G6V &  PMS, Var, IR, X \\
89.54925 & -35.01375 & 46 & 1.24 & $\mathrm{8^T}$ & Octans &  & 0.69 & 0.45 & -0.035 &  G9V &  X, PMS, Var, IR \\
168.35929 & -45.39522 & 67 & 2.50 & $\mathrm{8^T}$ & TW Hya &  &  & 0.92 & 0.108 &  M0.5 &  T Tau, Var, IR, X \\
170.27292 & -38.75453 & 34 & 2.08 & $\mathrm{8^T}$ & TW Hya & 6.6 & 1.45 & 0.95 & 0.096 &  M1Ve &  T Tau, Var, IR, X, Star \\
296.76612 & -78.96197 & 20 & 1.39 & $\mathrm{8^T}$ & Octans &  & 0.79 & 0.53 & -0.004 &  G8V &  PMS, Var, IR, X \\
179.92613 & -76.02392 & 22 & 1.16 & $\mathrm{9^T}$ & Eta Cha & 7.3 & 1.13 & 0.84 & -0.035 &  K4Ve &  T Tau, X, Var, IR \\
189.83850 & -75.04422 & 61 & 1.62 & $\mathrm{9^T}$ & Eta Cha &  & 0.98 & 0.66 & -0.026 &  K3Ve &  T Tau, X, Var, IR \\
194.60667 & -70.48031 & 125 & 1.98 & $\mathrm{9^T}$ & Eta Cha &  & 0.82 & 0.64 & -0.038 &  K0Ve &  PMS, X, Var, IR \\
200.53137 & -69.63672 & 25 & 3.24 & $\mathrm{9^T}$ & Eta Cha &  &  & 0.98 & 0.394 &  K1Ve &  Em, PMS, T Tau, Var, IR, X, Star \\
75.19646 & -57.25711 & 56 & 1.65 & $\mathrm{10^T}$ & Beta Pic & 7.3 & 1.19 & 0.85 & 0.036 &  M0Ve &  PMS, Var, IR, X \\
175.36513 & -73.78417 & 24 & 1.38 & $\mathrm{10^T}$ & Cha-Near &  & 0.76 & 0.53 & -0.019 &  G5 &  IR \\
281.71900 & -62.17686 & 64 & 1.80 & $\mathrm{10^T}$ & Beta Pic &  & 1.44 & 0.89 & 0.043 &  M1Ve &  Var, IR, X, PMS \\
320.20812 & -53.03417 & 56 & 1.34 & $\mathrm{10^W}$ & Tuc/Hor &  &  & 0.48 & -0.057 &  G7V &  PMS, Var, IR, UV, X \\
10.58471 & -77.79436 & 92 & 1.49 & $\mathrm{30^T}$ & Tuc/Hor &  & 0.99 & 0.69 & -0.028 &  K3Ve &  PMS, Var, IR, X \\
28.06092 & -52.32589 & 26 & 1.53 & $\mathrm{30^T}$ & Columba & 6.5 & 0.88 & 0.65 & -0.041 &  K2V(e) &  Var, IR, X, PMS \\
39.21542 & -52.05103 & 92 & 2.20 & $\mathrm{30^T}$ & Tuc/Hor &  & 1.42 & 0.92 & 0.103 &  M2Ve &  Var, X, Fl, IR \\
40.63750 & -57.66017 & 88 & 1.63 & $\mathrm{30^T}$ & Tuc/Hor &  & 1.19 & 0.78 & -0.024 &  K5Ve &  Var, X, PMS, IR \\
52.70458 & -45.93261 & 112 & 1.43 & $\mathrm{30^T}$ & Tuc/Hor &  & 1.01 & 0.67 & -0.028 &  K3V &  X, PMS, Var, IR \\
52.98183 & -43.98711 & 59 & 1.72 & $\mathrm{30^T}$ & Tuc/Hor & 7.4 & 1.22 & 0.83 & 0.008 &  K6Ve &  PMS, Var, IR, X \\
63.59404 & -38.31714 & 79 & 0.91 & $\mathrm{30^T}$ & Columba & 4.8 & 0.93 & 0.32 & -0.023 &  G3V &  PMS, Var, IR, X \\
65.29304 & -24.53919 & 73 & 1.32 & $\mathrm{30^T}$ & Columba &  &  & 0.40 & -0.023 &  G2V &  PMS, Var, IR, X \\
72.97312 & -46.78706 & 84 & 1.20 & $\mathrm{30^T}$ & Columba & 5.1 & 0.58 & 0.38 & -0.004 &  G5V &  PMS, Var, IR, X \\
73.20633 & -19.91711 & 69 & 1.48 & $\mathrm{30^T}$ & Tuc/Hor & 6.5 & 0.88 & 0.72 & -0.037 &  &  Var, IR, X \\
74.64904 & -15.62517 & 33 & 1.45 & $\mathrm{30^T}$ & Columba &  & 0.78 & 0.46 & -0.133 &  &  IR \\
74.70217 & -8.72772 & 25 & 1.68 & $\mathrm{30^T}$ & Columba &  & 0.71 & 0.51 & -0.033 &  &  IR, X \\
74.88346 & -19.29489 & 68 & 1.33 & $\mathrm{30^T}$ & Tuc/Hor &  & 1.04 & 0.70 & -0.060 &  &  Var \\
75.21608 & -41.01850 & 94 & 1.23 & $\mathrm{30^T}$ & Columba &  &  & 0.38 & -0.031 &  G5V &  IR, X, PMS \\
85.64267 & -34.26172 & 33 & 1.45 & $\mathrm{30^T}$ & Tuc/Hor &  & 0.82 & 0.54 & -0.042 &  &  Star \\
86.31767 & -38.61364 & 59 & 1.51 & $\mathrm{30^T}$ & Columba &  & 0.75 & 0.48 & -0.025 &  G9V &  Var, IR, X \\
96.52879 & -41.04828 & 101 & 1.16 & $\mathrm{30^T}$ & Columba & 5.8 & 0.68 & 0.51 & -0.037 &  K0V &  PMS, Var, IR, X \\
97.02533 & -48.44797 & 75 & 1.56 & $\mathrm{30^T}$ & Columba &  & 0.77 & 0.53 & -0.046 &  G9V &  Var, X, PMS, IR \\
110.34879 & -57.34361 & 55 & 1.37 & $\mathrm{30^T}$ & Carina & 5.8 & 0.72 & 0.51 & -0.019 &  K0V &  Var, X, PMS, IR \\
132.52254 & -75.91058 & 54 & 1.40 & $\mathrm{30^T}$ & Carina &  & 0.70 & 0.55 & -0.049 &  G9V &  T Tau, Var, X, IR \\
326.12550 & -60.97750 & 66 & 1.94 & $\mathrm{30^T}$ & Tuc/Hor & 7.2 & 1.35 & 0.88 & 0.026 &  M0Ve &  PMS, Var, IR, X \\
351.54454 & -73.39719 & 37 & 1.83 & $\mathrm{30^T}$ & Tuc/Hor & 6.9 & 1.43 & 0.90 & 0.058 &  M0Ve &  PMS, Var, IR, X \\
156.40950 & -64.68486 & 53 & 0.32 & $\mathrm{32^D}$ & IC 2602$^\mathrm{C,\,D}$ &  & 0.80 & 0.54 & -0.060 &  &  Star \\
159.57371 & -64.13514 & 47 & 0.69 & $\mathrm{32^D}$ & IC 2602$^\mathrm{C,\,D}$ &  &  & 0.38 & -0.032 &  G5 &  Cl, IR, X \\
159.57371 & -64.13514 & 107 & 0.68 & $\mathrm{32^D}$ & IC 2602$^\mathrm{C,\,D}$ &  &  & 0.38 & -0.032 &  G5 &  Cl, IR, X \\
160.00017 & -63.25308 & 58 & 0.72 & $\mathrm{32^D}$ & IC 2602$^\mathrm{C,\,D}$ &  & 0.68 & 0.36 & -0.040 &  &  IR, X \\
160.00017 & -63.25308 & 93 & 0.73 & $\mathrm{32^D}$ & IC 2602$^\mathrm{C,\,D}$ &  & 0.68 & 0.36 & -0.040 &  &  IR, X \\ 
116.85821 & -49.04753 & 50 & 0.86 & $\mathrm{40^T}$ & Argus &  & 0.68 & 0.50 & -0.036 &  G7V &  PMS, Var, IR \\
117.20762 & -43.45156 & 66 & 1.94 & $\mathrm{40^T}$ & Argus &  & 1.02 & 0.73 & 0.006 &  K4Ve &  PMS, Var, IR, X \\
118.48133 & -57.16897 & 33 & 1.11 & $\mathrm{40^T}$ & Argus &  & 0.81 & 0.59 & -0.032 &  K0V &  X, PMS, Var, IR \\
128.57567 & -52.26603 & 27 & 1.27 & $\mathrm{40^T}$ & Argus &  & 0.95 & 0.57 & -0.052 &  &  \\
128.93204 & -53.35564 & 27 & 1.70 & $\mathrm{40^T}$ & Argus$^\mathrm{C,\,D}$, IC 2391$^\mathrm{C,\,D}$ &  & 0.92 & 0.66 & -0.011 &  &  Star \\
129.22913 & -53.14286 & 50 & 1.18 & $\mathrm{40^T}$ & Argus &  & 0.80 & 0.50 & -0.056 &  &  Cl, Var \\
130.06771 & -52.94144 & 38 & 1.84 & $\mathrm{40^T}$ & Argus &  & 0.72 & 0.57 & -0.018 &  G9 &  Cl, Var, X \\
130.75171 & -53.90211 & 55 & 1.06 & $\mathrm{40^T}$ & Argus$^\mathrm{C,\,D}$, IC 2391$^\mathrm{C,\,D}$ &  & 0.69 & 0.50 & -0.005 &  &  IR \\
135.51642 & -58.14717 & 50 & 1.50 & $\mathrm{40^T}$ & Argus &  &  & 0.54 & -0.035 &  G8V &  X, PMS, Var \\
146.83275 & -40.05272 & 74 & 1.40 & $\mathrm{40^T}$ & Argus & 5.7 & 0.73 & 0.53 & -0.033 &  K0V &  PMS, IR, X \\
301.84904 & -51.79086 & 67 & 1.80 & $\mathrm{40^T}$ & Argus &  & 1.23 & 0.77 & -0.028 &  K6Ve &  PMS, Var, IR, X \\
130.75171 & -53.90211 & 23 & 0.88 & $\mathrm{45^D}$ & IC 2391$^\mathrm{C,\,D}$ &  & 0.69 & 0.50 & -0.005 &  &  IR \\
135.12738 & -61.57603 & 60 & 1.06 & $\mathrm{60^D}$ & Platais 8$^\mathrm{C,\,D}$ &  & 0.59 & 0.50 & -0.021 &  &  IR \\
0.79500 & -30.18025 & 46 & 0.46 & $\mathrm{62^D}$ & Blanco 1$^\mathrm{C,\,D}$ &  &  & 0.31 & -0.023 &  &  Cl, IR \\
1.16329 & -32.11753 & 65 & 0.18 & $\mathrm{62^D}$ & Blanco 1$^\mathrm{C,\,D}$ &  &  & 0.50 & -0.058 &  &  Star \\
1.51412 & -32.15508 & 62 & 0.36 & $\mathrm{62^D}$ & Blanco 1$^\mathrm{C,\,D}$ & 6.2 & 0.81 & 0.55 & -0.051 &  &  Star \\
1.64696 & -32.80175 & 38 & 0.32 & $\mathrm{62^D}$ & Blanco 1$^\mathrm{C,\,D}$ & 4.8 & 0.62 & 0.37 & -0.019 &  &  IR \\
74.34304 & -9.13317 & 75 & 1.28 & $\mathrm{70^T}$ & AB Dor & 5.2 & 0.73 & 0.40 & -0.051 &  G5 &  IR \\
75.62679 & -39.98694 & 39 & 1.31 & $\mathrm{70^T}$ & AB Dor & 7.1 & 1.04 & 0.68 & -0.046 &  K4V &  X, PMS, Var, IR \\
76.61533 & -15.82506 & 104 & 0.63 & $\mathrm{70^T}$ & AB Dor &  & 0.53 & 0.29 & -0.034 &  F8V &  Var, IR, X \\
90.59129 & -13.92569 & 59 & 1.03 & $\mathrm{70^T}$ & AB Dor &  & 1.07 & 0.67 & -0.043 &  &  IR, X \\
92.14108 & -34.04861 & 37 & 1.44 & $\mathrm{70^T}$ & AB Dor &  & 0.75 & 0.52 & -0.017 &  G9Ve &  PMS, Var, IR, X \\
100.32708 & -38.34333 & 59 & 2.25 & $\mathrm{70^T}$ & AB Dor &  & 0.92 & 0.66 & -0.019 &  K2Ve &  PMS, Var, IR, X \\
101.97237 & -57.22556 & 29 & 0.98 & $\mathrm{70^T}$ & AB Dor & 7.3 & 0.99 & 0.75 & -0.048 &  K4V &  Var, X, IR \\
112.74804 & -84.32433 & 71 & 1.33 & $\mathrm{70^T}$ & AB Dor &  & 0.51 & 0.50 & -0.025 &  G9V &  PMS, Var, IR, X \\
151.85475 & -46.36378 & 36 & 0.95 & $\mathrm{70^T}$ & AB Dor &  & 1.09 & 0.70 & -0.064 &  K4V &  PMS, IR, X \\
287.74096 & -60.27228 & 124 & 0.66 & $\mathrm{70^T}$ & AB Dor &  &  & 0.32 & -0.013 &  G1V &  PMS, Var, IR, X \\
318.27200 & -17.48686 & 49 & 1.11 & $\mathrm{70^T}$ & AB Dor & 7.5 & 1.16 & 0.77 & -0.075 &  K6Ve &  PMS, Var, IR, X \\
53.50754 & 24.88089 & 53 & 0.54 & $\mathrm{135^D}$ & Pleiades$^\mathrm{C}$ & 5.4 & 0.71 & 0.40 & -0.046 &  &  Cl, IR \\
53.88204 & 22.82358 & 71 & 0.22 & $\mathrm{135^D}$ & Pleiades$^\mathrm{C}$ &  &  & 0.29 & -0.051 &  &  Cl, IR \\
54.59408 & 22.49969 & 68 & 0.55 & $\mathrm{135^D}$ & Pleiades$^\mathrm{C}$ &  & 0.56 & 0.32 & -0.050 &  F8 &  Cl, IR \\
54.73696 & 24.56981 & 59 & 0.44 & $\mathrm{135^D}$ & Pleiades$^\mathrm{C}$ &  & 0.60 & 0.32 & -0.043 &  &  Cl, IR \\
54.80617 & 24.46653 & 65 & 0.54 & $\mathrm{135^D}$ & Pleiades$^\mathrm{C,\,D}$ &  & 0.60 & 0.33 & -0.021 &  &  Cl, IR \\
54.86583 & 23.89500 & 41 & 0.66 & $\mathrm{135^D}$ & Pleiades$^\mathrm{C}$ &  & 0.85 & 0.48 & -0.057 &  &  Cl \\
54.92158 & 23.29089 & 72 & 0.43 & $\mathrm{135^D}$ & Pleiades$^\mathrm{C,\,D}$ &  & 0.45 & 0.29 & -0.028 &  G &  Cl, IR \\
55.12804 & 24.48731 & 27 & 0.83 & $\mathrm{135^D}$ & Pleiades$^\mathrm{C}$ &  & 0.86 & 0.53 & -0.017 &  &  Cl, IR \\
55.14317 & 23.68261 & 53 & 0.71 & $\mathrm{135^D}$ & Pleiades$^\mathrm{C}$ &  & 0.73 & 0.44 & -0.043 &  &  Cl, IR \\
55.36592 & 23.70836 & 49 & 0.84 & $\mathrm{135^D}$ & Pleiades$^\mathrm{C}$ &  & 0.75 & 0.45 & -0.062 &  &  Cl \\
55.40067 & 25.61931 & 80 & 0.48 & $\mathrm{135^D}$ & Pleiades$^\mathrm{C,\,D}$ &  & 0.58 & 0.37 & -0.021 &  &  Cl, IR \\
56.72400 & 23.58356 & 52 & 0.44 & $\mathrm{135^D}$ & Pleiades$^\mathrm{C}$ &  & 0.57 & 0.35 & -0.047 &  G0V &  Cl, IR \\
56.72400 & 23.58356 & 78 & 0.51 & $\mathrm{135^D}$ & Pleiades$^\mathrm{C}$ &  & 0.57 & 0.35 & -0.047 &  G0V &  Cl, IR \\
57.35717 & 24.93764 & 33 & 0.74 & $\mathrm{135^D}$ & Pleiades$^\mathrm{C,\,D}$ &  & 0.87 & 0.48 & -0.057 &  G2 &  Cl, IR, X \\
57.35717 & 24.93764 & 23 & 0.90 & $\mathrm{135^D}$ & Pleiades$^\mathrm{C}$ &  & 0.87 & 0.48 & -0.057 &  G2 &  Cl, IR, X \\
57.58875 & 23.09639 & 26 & 0.48 & $\mathrm{135^D}$ & Pleiades$^\mathrm{C}$ &  & 0.91 & 0.48 & -0.065 &  &  Cl, IR \\
57.58875 & 23.09639 & 57 & 0.49 & $\mathrm{135^D}$ & Pleiades$^\mathrm{C}$ &  & 0.91 & 0.48 & -0.065 &  &  Cl, IR \\
57.92533 & 21.66836 & 38 & 0.63 & $\mathrm{135^D}$ & Pleiades$^\mathrm{C}$ &  & 0.83 & 0.45 & -0.067 &  &  Cl \\
58.00933 & 24.66331 & 20 & 2.09 & $\mathrm{135^D}$ & Pleiades$^\mathrm{C}$ &  & 1.07 & 0.73 & 0.008 &  K3V &  Cl, Em, T Tau, Var, Fl, IR, X \\
327.59629 & -16.19647 & 132 & 0.47 & $\mathrm{150^I}$ & -4.33 &  & 0.65 & 0.40 & -0.073 &  G7V &  IR \\
209.06946 & -62.35256 & 30 & 0.20 & $\mathrm{169^D}$ & Platais 12$^\mathrm{C,\,D}$ &  & 0.60 & 0.34 & -0.041 &  &  IR \\
65.88471 & 14.67047 & 93 & 0.67 & $\mathrm{380^I}$ & Hyades &  & 0.73 & 0.47 & -0.034 &  G5 &  Cl, X, IR \\
67.49054 & 16.67283 & 131 & 0.56 & $\mathrm{400^I}$ & Hyades & 6.6 & 0.79 & 0.48 & -0.062 &  K1V &  Cl, X, IR \\
0.16796 & -69.67597 & 121 & 0.72 & $\mathrm{512^A}$ & -4.47 & 6.6 & 0.87 & 0.55 & -0.033 &  K1V &  IR \\
50.58333 & -36.63717 & 72 & 1.28 & $\mathrm{524^K}$ & Alessi 13$^\mathrm{C}$ & 5.0 & 0.73 & 0.41 & -0.039 &  G5 &  IR \\
53.08567 & -38.74736 & 23 & 0.05 & $\mathrm{524^K}$ & Alessi 13$^\mathrm{C}$ & 6.2 & 0.74 & 0.47 & -0.059 &  &  IR \\
67.01850 & 13.86794 & 55 & 0.64 & $\mathrm{630^I}$ & Hyades &  & 0.72 & 0.50 & -0.034 &  G5 &  Cl, IR \\
66.05325 & 16.37892 & 67 & 0.22 & $\mathrm{787^D}$ & Hyades &  & 0.56 & 0.28 & -0.021 &  G0V &  Cl, X, IR \\
66.85558 & 14.26067 & 83 & 0.70 & $\mathrm{787^D}$ & Hyades &  & 1.04 & 0.66 & -0.051 &  K2 &  Cl, IR \\
66.94600 & 14.41775 & 46 & 0.51 & $\mathrm{787^D}$ & Hyades &  & 0.70 & 0.55 & -0.040 &  K0 &  Cl, X, IR \\
67.04529 & 16.47094 & 79 & 0.03 & $\mathrm{787^D}$ & Hyades &  & 1.37 & 0.80 & -0.068 &  K5 &  Cl, IR \\
68.40817 & 16.76250 & 99 & 0.71 & $\mathrm{787^D}$ & Hyades & 6.8 & 0.87 & 0.49 & -0.042 &  G5 &  Cl, hPM, X, IR \\
68.63408 & 15.82756 & 116 & 0.68 & $\mathrm{787^D}$ & Hyades &  & 0.80 & 0.51 & -0.057 &  K0 &  Cl, IR \\
71.89725 & 14.88908 & 48 & 0.30 & $\mathrm{787^D}$ & Hyades & 8.4 & 1.34 & 0.74 & -0.067 &  M0V &  Cl, Var \\
72.21533 & 15.94756 & 68 & 0.78 & $\mathrm{787^D}$ & Hyades &  &  & 0.50 & -0.060 &  &  IR \\
72.50296 & 16.41206 & 76 & 0.02 & $\mathrm{787^D}$ & Hyades &  & 1.10 & 0.65 & -0.080 &  K0 &  Cl, IR \\
72.95496 & 17.27372 & 37 & 0.37 & $\mathrm{787^D}$ & Hyades &  &  & 0.74 & -0.079 &  K5 &  hPM \\
74.25283 & 13.91236 & 104 & 0.23 & $\mathrm{787^D}$ & Hyades &  & 1.13 & 0.64 & -0.092 &  &  Cl \\
75.40696 & 12.41639 & 53 & 0.36 & $\mathrm{787^D}$ & Hyades & 8.1 & 1.34 & 0.85 & 0.013 &  &  Star \\
39.03250 & -28.21833 & 82 & -0.00 & $\mathrm{880^I}$ & -4.67 &  & 0.76 & 0.43 & -0.067 &  G9V &  IR \\
74.45629 & 14.00219 & 61 & 0.48 & $\mathrm{970^I}$ & Hyades & -4.57 & 0.65 & 0.33 & -0.076 &  G5 &  Cl, IR \\
69.73879 & 14.10558 & 119 & 0.40 & $350^\mathrm{I}, 1122^\mathrm{W}$ & Hyades &  &  & 0.37 & -0.046 &  G5 &  Cl, IR \\
47.17717 & -35.43289 & 213 & -0.02 & $\mathrm{1220^I}$ & -4.61 & 5.7 & 0.56 & 0.41 & 0.213 &  G3V &  IR \\
244.22125 & -49.85675 & 114 & 0.33 & $\mathrm{1230^A}$ & -4.67 &  &  & 0.45 & -0.032 &  G8IV-V &  IR \\
277.82900 & -18.90881 & 133 & 0.36 & $1260^\mathrm{I}, 1621^\mathrm{W}$ & -4.71, -4.65 &  &  & 0.58 & 0.350 &  K2V &  Cl, hPM, Var, IR \\
36.86792 & -27.63500 & 41 & 0.06 & $\mathrm{1940^I}$ & -4.70 &  &  & 0.51 & -0.062 &  K1V &  hPM, IR \\
5.64492 & -73.03236 & 123 & 0.17 & $\mathrm{2290^A}$ & -4.73 &  & 0.70 & 0.41 & -0.060 &  G8V &  IR \\
17.24133 & -30.92925 & 71 & 0.15 & $\mathrm{2570^A}$ & -4.75 &  &  & 0.47 & -0.073 &  G8/K0V &  hPM, IR \\
25.65417 & -58.41317 & 150 & -0.00 & $\mathrm{3019^A}$ & -4.79 &  & 0.79 & 0.42 & -0.034 &  G6V &  IR \\
30.19675 & -80.53269 & 36 & 0.09 & $\mathrm{3548^A}$ & -4.82 &  &  & 0.45 & -0.047 &  K1V &  IR \\
279.72250 & -21.05186 & 238 & 0.08 & $5360^\mathrm{I}, 3890^\mathrm{W}$ & -4.93, -4.90 &  &  & 0.56 & 0.352 &  G6V &  IR, UV \\
1.24463 & -70.21244 & 82 & 0.10 & $\mathrm{4073^A}$ & -4.88 & 5.8 & 0.66 & 0.38 & -0.036 &  G5V &  IR \\
180.18521 & -10.44603 & 184 & 0.22 & $7350^\mathrm{I}, 4168^\mathrm{W}$ & -4.89, -4.92 &  &  & 0.53 &  &  G8IV &  hPM, IR, UV \\
183.46604 & -69.06050 & 119 & 0.03 & $\mathrm{5370^A}$ & -4.97 &  &  & 0.56 & -0.040 &  K2V &  hPM, IR \\
26.90958 & -26.75000 & 45 & -0.06 & $6860^\mathrm{I}, 5495^\mathrm{W}$ & -5.04, -5.00 &  &  & 0.50 & -0.061 &  G9V &  hPM, IR \\
168.00483 & -26.13664 & 123 & 0.17 & $7470^\mathrm{I}, 5495^\mathrm{W}$ & -5.01, -5.00 &  &  & 0.48 & 0.163 &  G8.5V &  hPM, IR \\
193.84079 & -4.50417 & 185 & 0.03 & $\mathrm{5940^I}$ & -4.97 & 6.5 & 0.70 & 0.46 & 0.127 &  G5 &  hPM, IR \\
260.71371 & -2.38817 & 237 & -0.08 & $5810^\mathrm{I}, 6309^\mathrm{W}$ & -5.03, -5.04 &  &  & 0.45 & 0.436 &  G5IV &  IR, UV \\
265.93437 & -3.91786 & 154 & 0.03 & $\mathrm{6456^A}$ & -4.99 &  &  & 0.40 & 0.024 &  G5 &  hPM, IR \\
32.53333 & -31.06972 & 38 & 0.09 & $\mathrm{6918^A}$ & -5.02 &  &  & 0.42 & -0.085 &  K0V &  hPM, IR \\
209.11283 & -16.19622 & 78 & -0.14 & $\mathrm{7000^I}$ & -5.02 &  &  & 0.44 & -0.071 &  G9V &  hPM, IR \\
189.02742 & -42.85231 & 31 & 0.45 & $\mathrm{8128^A}$ & -5.00 &  &  & 0.37 & -0.022 &  G5V &  IR \\
318.36537 & -49.79458 & 120 & 0.07 & $\mathrm{9120^A}$ & -5.15 &  & 0.52 & 0.41 & -0.054 &  G6IV &  IR \\
16.99858 & 1.99306 & 111 & -0.16 & $\mathrm{10964^W}$ & -4.92 &  &  & 0.60 & 0.345 &  K0IV &  hPM, IR \\
\label{tab.reference_stars2}
\end{longtable}
\end{center}
\clearpage
\end{landscape}



\end{document}